\documentclass[preprint,english,showpacs,11pt,floatfix]{revtex4}

\usepackage{babel,amsmath,amssymb,dcolumn}
\usepackage[dvips]{graphics}
\usepackage{hyperref}

\usepackage{amsfonts}
\usepackage{amssymb}
\usepackage[dvips]{graphicx}
\usepackage{latexsym}
\usepackage{subfigure}

\usepackage{dcolumn}
\usepackage{bm}

\newcommand{\be}{\begin{equation}}
\newcommand{\ee}{\end{equation}}

\begin{document}

\title{Testing the viability of the interacting holographic dark energy model
by using combined observational constraints}

\author{Chang Feng}
\author{Bin Wang}
\email{wangb@fudan.edu.cn}
\affiliation{Department of Physics, Fudan University, 200433
Shanghai}
\author{Yungui Gong}
\email{yungui_gong@baylor.edu} \affiliation{School of Physical
Science and Technology, Southwest University, Chongqing 400715,
China}
\author{Ru-Keng Su}
\email{rksu@fudan.ac.cn} \affiliation{China Center of Advanced
Science and Technology (World Laboratory) P.O. Box 8730, 100080
Beijing and Department of Physics, Fudan University, Shanghai
200433, China}

\begin{abstract}
Using the data coming from the new 182 Gold type Ia supernova
samples, the shift parameter of the Cosmic Microwave Background
given by the three-year Wilkinson Microwave Anisotropy Probe
observations, and the baryon acoustic oscillation measurement from
the Sloan Digital Sky Survey, $H(z)$ and lookback time measurements,
we have performed a statistical joint analysis of the interacting
holographic dark energy model. Consistent parameter estimations show
us that the interacting holographic dark energy model is a viable
candidate to explain the observed acceleration of our universe.
\end{abstract}

\pacs{98.80.Cq; 98.80.-k}

\maketitle

The present continuous flow of cosmological data provides us day by
day a clearer picture that our universe is experiencing accelerated
expansion \cite{agr98}. In order to draw precise conclusions from
the available phenomenological perspectives on how fast the universe
is expanding at present, how long this speed up has lasted and how
the acceleration rate has changed over the recent past, we have to
overcome statistical uncertainties and possible theoretical biasing
in the tests used. This requires us to refine the existing tests and
devise new ones.

Recently Simon et al \cite{2} have published Hubble parameter data
extracted from differential ages of passively evolving galaxies. It
is interesting to use these data to constrain the evolution of the
universe. This is so because that they can provide consistent checks
and tight constraints on models when combined with other
cosmological tests, and also because the Hubble parameter is not
integrated over like that of the luminosity distance and it can give
better constraints on the cosmological parameters. Recently, Hubble
parameter data have been used to constrain several cosmological
models \cite{3,4}.

To reduce the degeneracy in viable candidate cosmological models
designed to explain the observed accelerated expansion, new
observables should be added to the usual ones. Recalling that the
test of cosmological models by the type Ia supernova (SN Ia) data is
a distance based method, it is of interest to look for tests based
on time-dependent observable. In \cite{6,7}, the age of an old high
redshift galaxy has been used to constrain the model. To overcome
the problem that the estimate of the age of a single galaxy maybe
affected by systematic errors, it is needed to consider a sample of
galaxies belonging to the same cluster. Recently, the age estimates
of around 160 galaxy clusters at six redshifts distributed in the
interval $0.10<z<1.27$ have been compiled by Capozziello et al
\cite{8}. Employing these data, one can take into account the
lookback time which was defined by Sandage \cite{5} as the
difference between the present age of the universe and its age when
a particular light ray at redshift $z$ was emitted. This quantity
can discriminate among different cosmological models. The lookback
time has been used as a test for some cosmological models
\cite{8,9}.

In this paper we will use the latest SN Ia data compiled by Riess et
al \cite{10}, the Cosmic Microwave Background (CMB) shift parameter
derived from the three-year Wilkinson Microwave Anisotropy Probe
(WMAP3) observations \cite{11}, the baryon acoustic oscillations
(BAO) measurement from the large-scale correlation function of the
Sloan Digital Sky Survey (SDSS) luminous red galaxies \cite{12} in
combination with the $H(z)$ data and the lookback time data to give
a complete investigation on the viability of the interacting
holographic dark energy model devised in \cite{13}. Recently, this
model has confronted the tests from the SN Ia data \cite{13}, the
age constraint and the small $l$ CMB spectrum constraint \cite{7}.
It has been argued that the interacting holographic dark energy
model can accommodate the transition of the dark energy equation of
state $w$ from $w>-1$ to $w<-1$ \cite{13,14}, as recently revealed
from extensive data analysis \cite{15}. With the interaction between
dark energy and dark matter introduced in \cite{13}, it has been
shown that the old astrophysical structures can be formed naturally
\cite{7} and the coincidence problem can be alleviated \cite{16,7}.
The thermodynamical properties of the universe with the interacting
holographic dark energy have also been studied \cite{17}. Very
recently, the combined constraint on the interacting holographic
dark energy model using the SN Ia data, the BAO measurement and the
shift parameter determined from the SDSS and WMAP3 has been reported
\cite{CC}. This paper aims to place combined new observational
constraints on this interacting holographic dark energy model by
including the Hubble parameter data and the lookback time data.
Different from the distance based test, the lookback time is a time
based method. Moreover, the Hubble parameter does not suffer the
integration effect in the luminosity distance. It is expected that
these new tests will further constrain the model.

Recently, inspired by the holographic hypothesis \cite{18}, a new
model has been put forward to explain the dark energy. The energy
density cannot exceed the mass of a black hole with the same size of
the universe $L$, thus we have $\rho_D=3c^2L^{-2}$, where $c$ is a
constant and the Planck mass $M_p$ has been taken unity. Choosing
$L$ as the future event horizon,
$R_h=a\int_{a}^{\infty}\frac{da}{Ha^2}$, we have $\rho_D=3c^2
R_h^{-2}$ as the dark energy density. As far as energy conservation
is concerned, we suppose that the interaction is described by the
(separately non conserving) equations
\begin{equation}
\dot{\rho_m}+3H\rho_m=Q
\end{equation}
\begin{equation}
\dot{\rho_D}+3H(1+\omega_D)=-Q
\end{equation}
where $Q$ is some interaction term. For the moment we take for
granted that the interaction is the one proposed on general grounds
in \cite{19}, which is $Q=3b^2H(\rho_m+\rho_D)$, where $b^2$ is the
second phenomenological constant indicating coupling between dark
energy and dark matter. Positive values of $b^2$ would correspond to
a transfer of energy from the dark energy to dark matter, while the
negative $b^2$ would imply a transfer of energy from the dark matter
to the dark energy \cite{p}. In view of the unknown nature of dark
matter and dark energy, we do not put any limit on the sign of $b^2$
at first and wait to determine it from the observational data.
Because of the interaction, neither dark energy nor dark matter
conserve whence they evolve separately. For the flat universe, using
the Friedmann equation $\Omega_D+\Omega_m=1$, where
$\Omega_D=\frac{\rho_D}{3H^2}$ and $\Omega_m=\frac{\rho_m}{3H^2}$,
the evolution behavior of the dark energy was obtained as \cite{13}:
\begin{equation}\label{eq3}
    \frac{\Omega'_D}{\Omega^2_D}=(1-\Omega_D)[\frac{2}{c\sqrt{\Omega_D}}
    +\frac{1}{\Omega_D}-\frac{3b^2}{\Omega_D(1-\Omega_D)}].
\end{equation}
The prime denotes the derivative with respect to $x=\ln a$. The
equation of state of dark energy was expressed as \cite{13}
\begin{equation}\label{eq4}
    \omega_D=-\frac{1}{3}-\frac{2\sqrt{\Omega_D}}{3c}-\frac{b^2}{\Omega_D}.
\end{equation}

By suitably choosing the coupling between dark energy and dark
matter, this model can accommodate the transition of the dark energy
equation of state from $\omega_D>-1$ to $\omega_D<-1$ \cite{13,14},
which is in agreement with the recent analysis of the SN Ia data
\cite{15}. The deceleration parameter has the form
\begin{equation}
q=\frac{1}{2}-\frac{3b^2}{2}-\frac{\Omega_D}{2}-\frac{\Omega_D^{3/2}}{c}.
\end{equation}

The evolution of the Hubble parameter can be written as
\begin{equation}
H(z)=H_0\exp{[\int_0^z\frac{1+q^{\prime}}{1+z^{\prime}}dz^{\prime}]}
\end{equation}

Next we constrain the interacting holographic dark energy model by
using the latest observational data, such as the gold SN Ia data,
the shift parameter and the BAO measurement from WMAP3 and SDSS, and
combining these observations with $H(z)$ data and lookback time
data.

\begin{figure}
  \subfigure[ ]{
    \label{fig:mini:subfig:a} 
    \begin{minipage}[b]{0.5\textwidth}
      \centering
      \includegraphics[width=8cm,height=8cm]{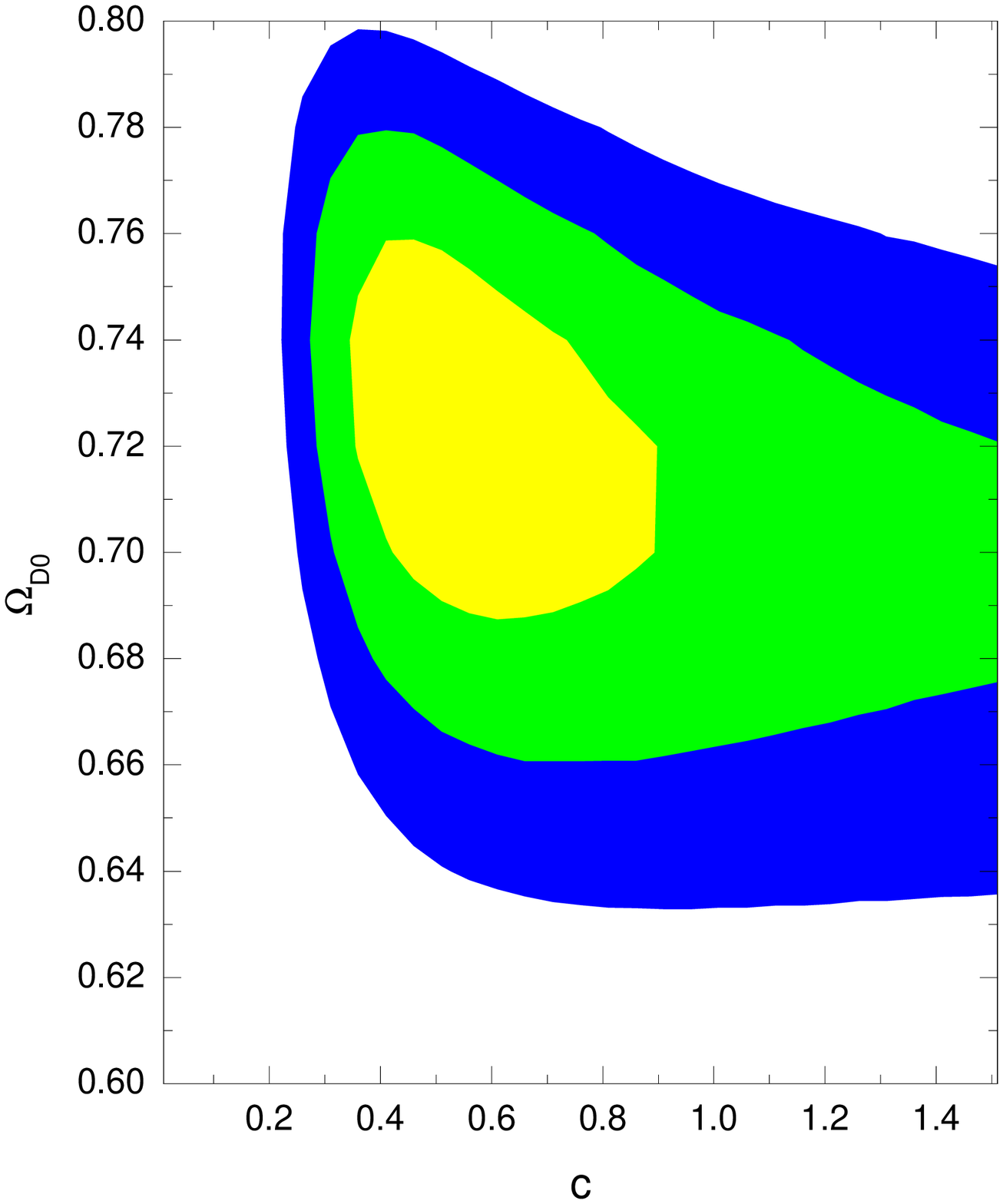}
    \end{minipage}}%
  \subfigure[]{
    \label{fig:mini:subfig:b} 
    \begin{minipage}[b]{0.5\textwidth}
      \centering
      \includegraphics[width=8cm,height=8cm]{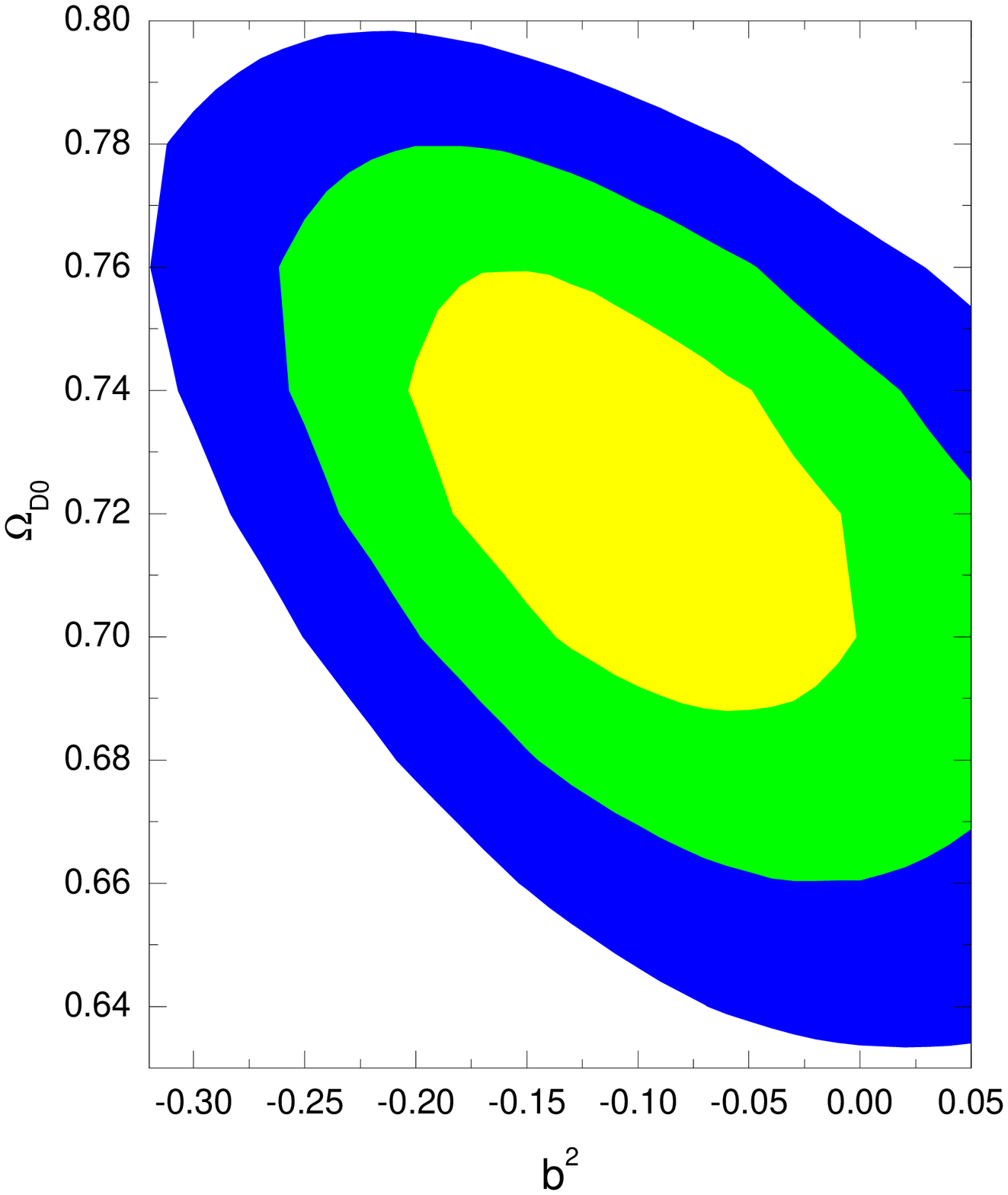}
    \end{minipage}}
    \caption{$(a)$The contours from the combination of
SN Ia, BAO in the interacting holographic dark energy model for $c$
and $\Omega_{D0}$ at $1\sigma$, $2\sigma$, $3\sigma$ confidence
level with $b^2=-0.10$. $(b)$The contours from the combination of SN
Ia, BAO for $b^2$ and $\Omega_{D0}$ at $1\sigma$, $2\sigma$,
$3\sigma$ confidence level with $c=0.53$.}
  \label{fig:mini:subfig} 
\end{figure}

\begin{figure}
  \subfigure[ ]{
    \label{fig:mini:subfig:a} 
    \begin{minipage}[b]{0.5\textwidth}
      \centering
      \includegraphics[width=8cm,height=8cm]{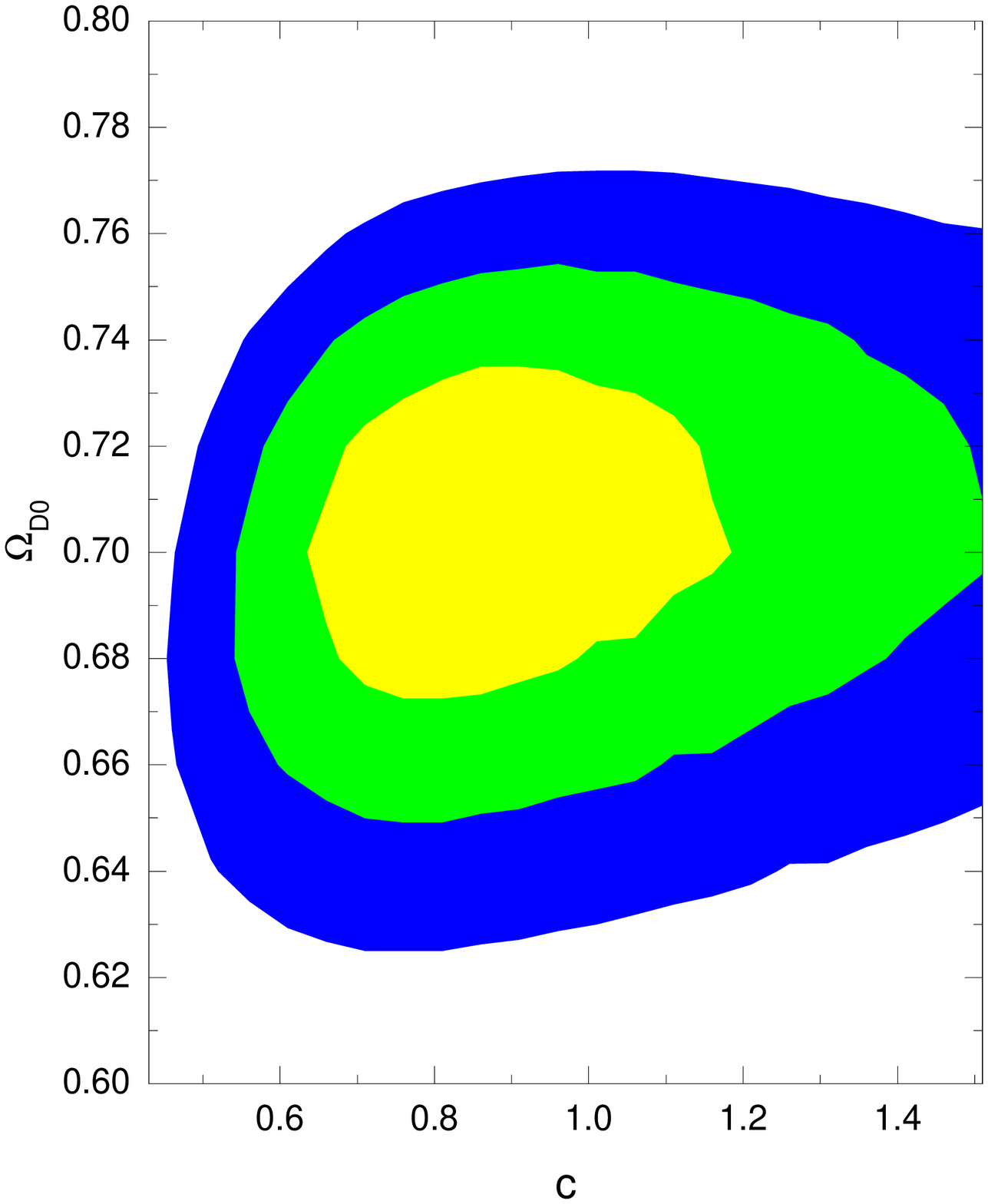}
    \end{minipage}}%
  \subfigure[]{
    \label{fig:mini:subfig:b} 
    \begin{minipage}[b]{0.5\textwidth}
      \centering
      \includegraphics[width=8cm,height=8cm]{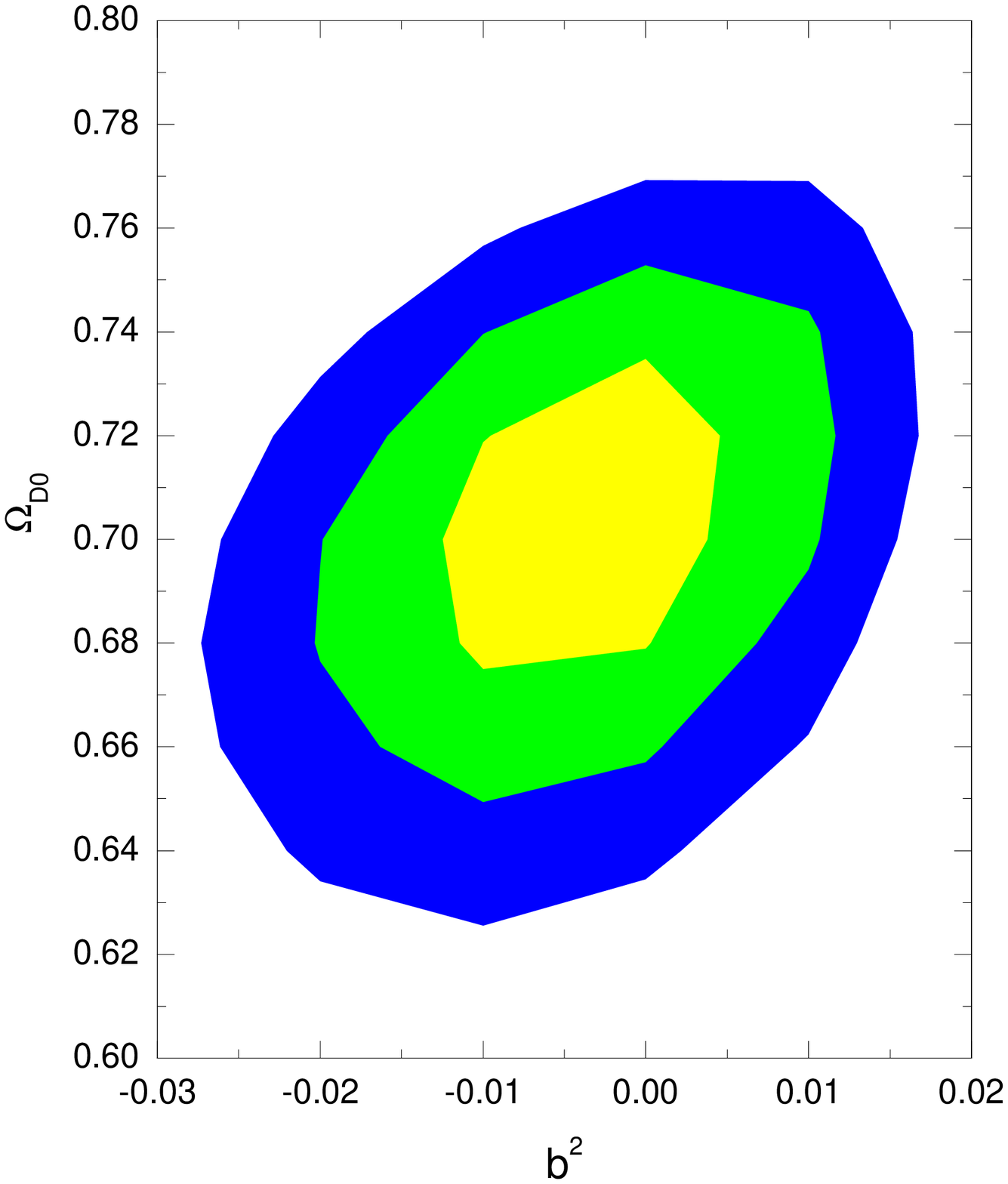}
    \end{minipage}}
    \caption{$(a)$The contours from the combination of
SN Ia, BAO, CMB in the interacting holographic dark energy model for
$c$ and $\Omega_{D0}$ at $1\sigma$, $2\sigma$, $3\sigma$ confidence
level with $b^2=-0.004$. $(b)$The contours from the combination of
SN Ia, BAO, CMB for $b^2$ and $\Omega_{D0}$ at $1\sigma$, $2\sigma$,
$3\sigma$ confidence level with $c=0.84$.}
  \label{fig:mini:subfig} 
\end{figure}

\begin{figure}
  \subfigure[ ]{
    \label{fig:mini:subfig:a} 
    \begin{minipage}[b]{0.5\textwidth}
      \centering
      \includegraphics[width=8cm,height=8cm]{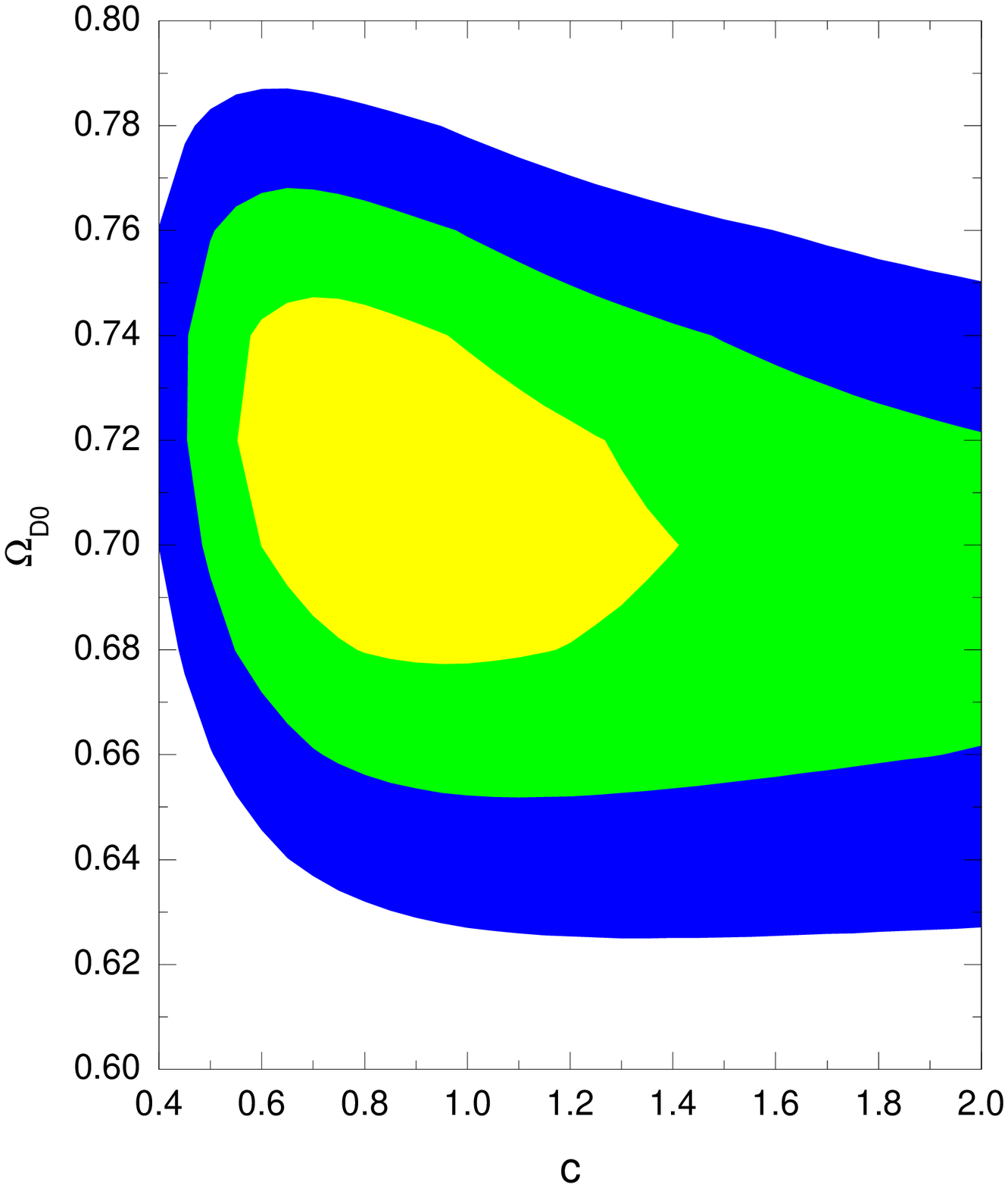}
    \end{minipage}}%
  \subfigure[]{
    \label{fig:mini:subfig:b} 
    \begin{minipage}[b]{0.5\textwidth}
      \centering
      \includegraphics[width=8cm,height=8cm]{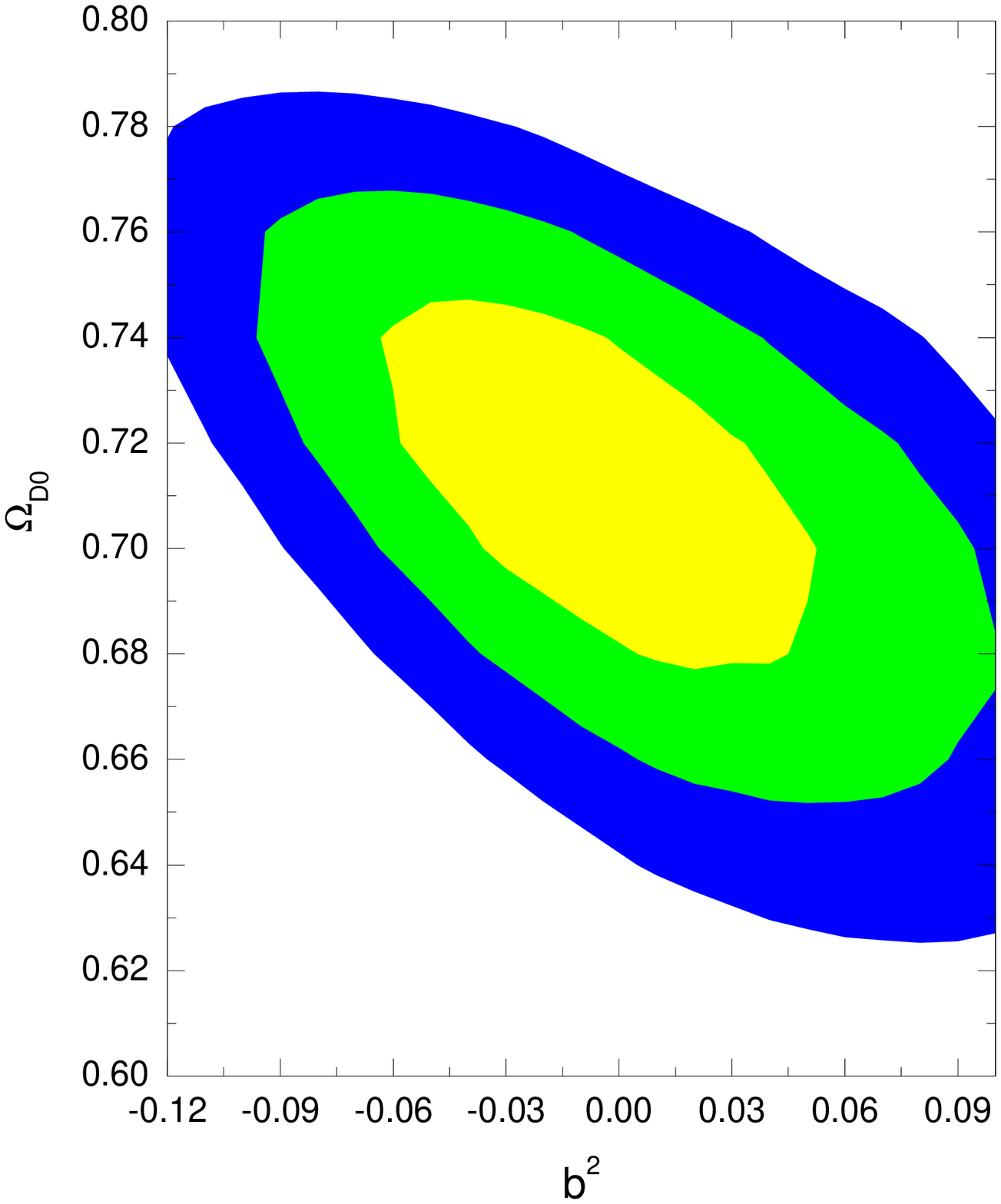}
    \end{minipage}}
    \caption{$(a)$The contours from the combination of
SN Ia, BAO, $H(z)$ in the interacting holographic dark energy model
for $c$ and $\Omega_{D0}$ at $1\sigma$, $2\sigma$, $3\sigma$
confidence level with $b^2=-0.005$. $(b)$The contours from the
combination of SN Ia, BAO, $H(z)$ for $b^2$ and $\Omega_{D0}$ at
$1\sigma$, $2\sigma$, $3\sigma$ confidence level with $c=0.82$. We
have employed $H_0=72km\cdot s^{-1}\cdot Mpc^{-1}$.}
  \label{fig:mini:subfig} 
\end{figure}

\begin{figure}
  \subfigure[ ]{
    \label{fig:mini:subfig:a} 
    \begin{minipage}[b]{0.5\textwidth}
      \centering
      \includegraphics[width=8cm,height=8cm]{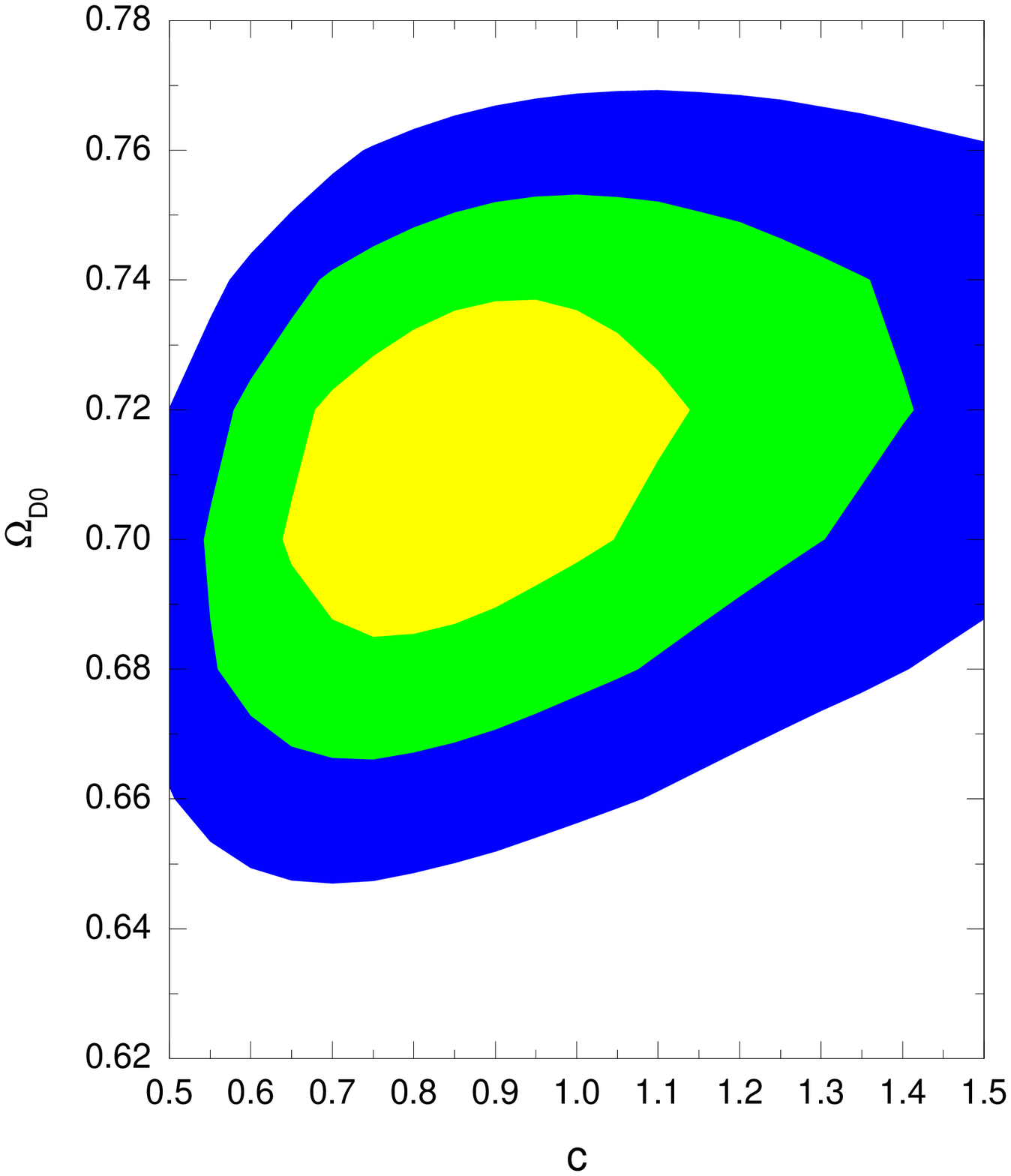}
    \end{minipage}}%
  \subfigure[]{
    \label{fig:mini:subfig:b} 
    \begin{minipage}[b]{0.5\textwidth}
      \centering
      \includegraphics[width=8cm,height=8cm]{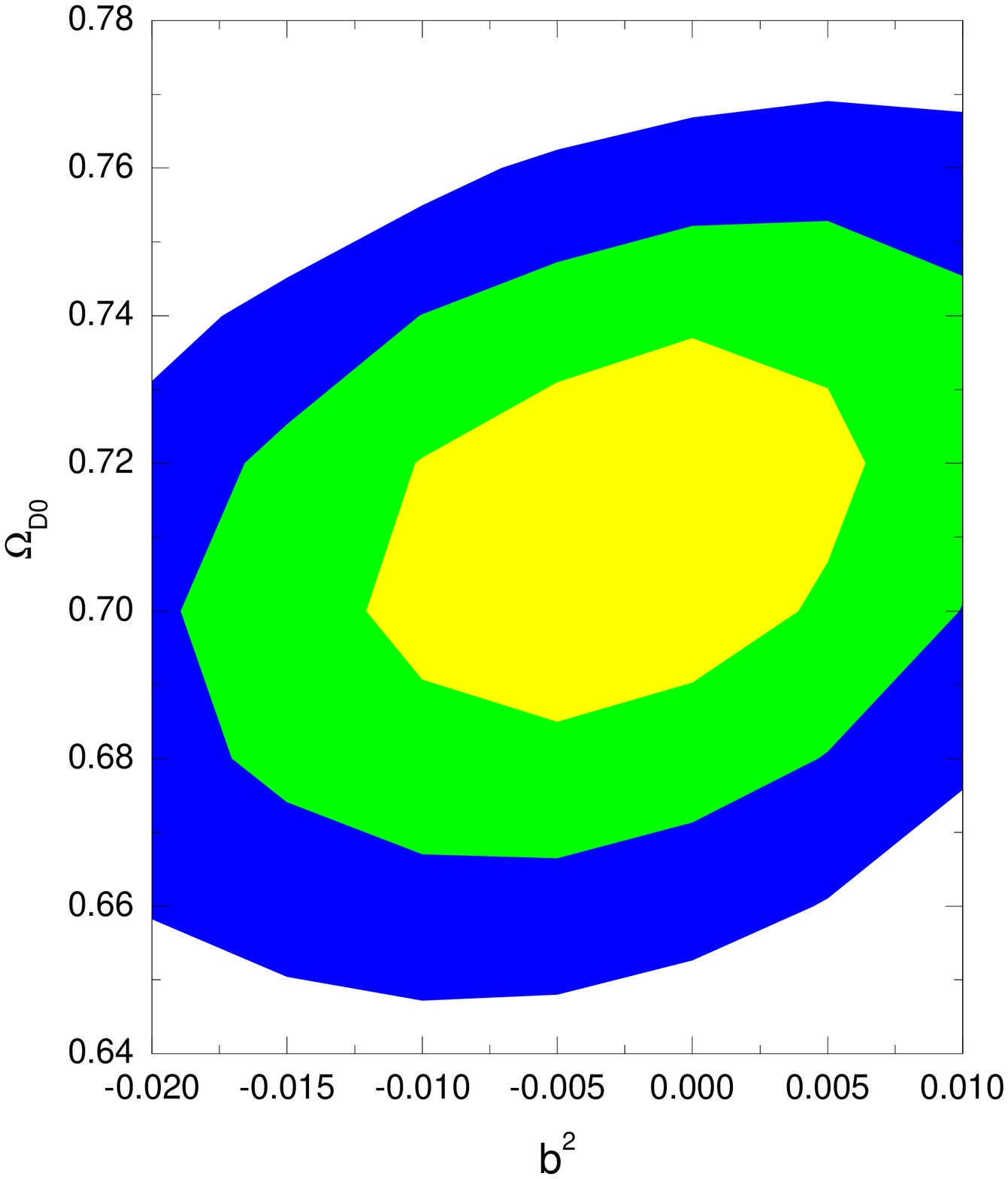}
    \end{minipage}}
    \caption{$(a)$The contours from the combination of
SN Ia, BAO, $H(z)$, CMB in the interacting holographic dark energy
model for $c$ and $\Omega_{D0}$ at $1\sigma$, $2\sigma$, $3\sigma$
confidence level with $b^2=-0.003$. $(b)$The contours from the
combination of SN Ia ,BAO, $H(z)$, CMB for $b^2$ and $\Omega_{D0}$
at $1\sigma$, $2\sigma$, $3\sigma$ confidence level with $c=0.84$.
We have employed $H_0=72km\cdot s^{-1}\cdot Mpc^{-1}$.}
  \label{fig:mini:subfig} 
\end{figure}

\begin{figure}
    \subfigure[]{
    \label{fig:mini:subfig:a} 
    \begin{minipage}[b]{0.5\textwidth}
      \centering
      \includegraphics[width=8cm,height=8cm]{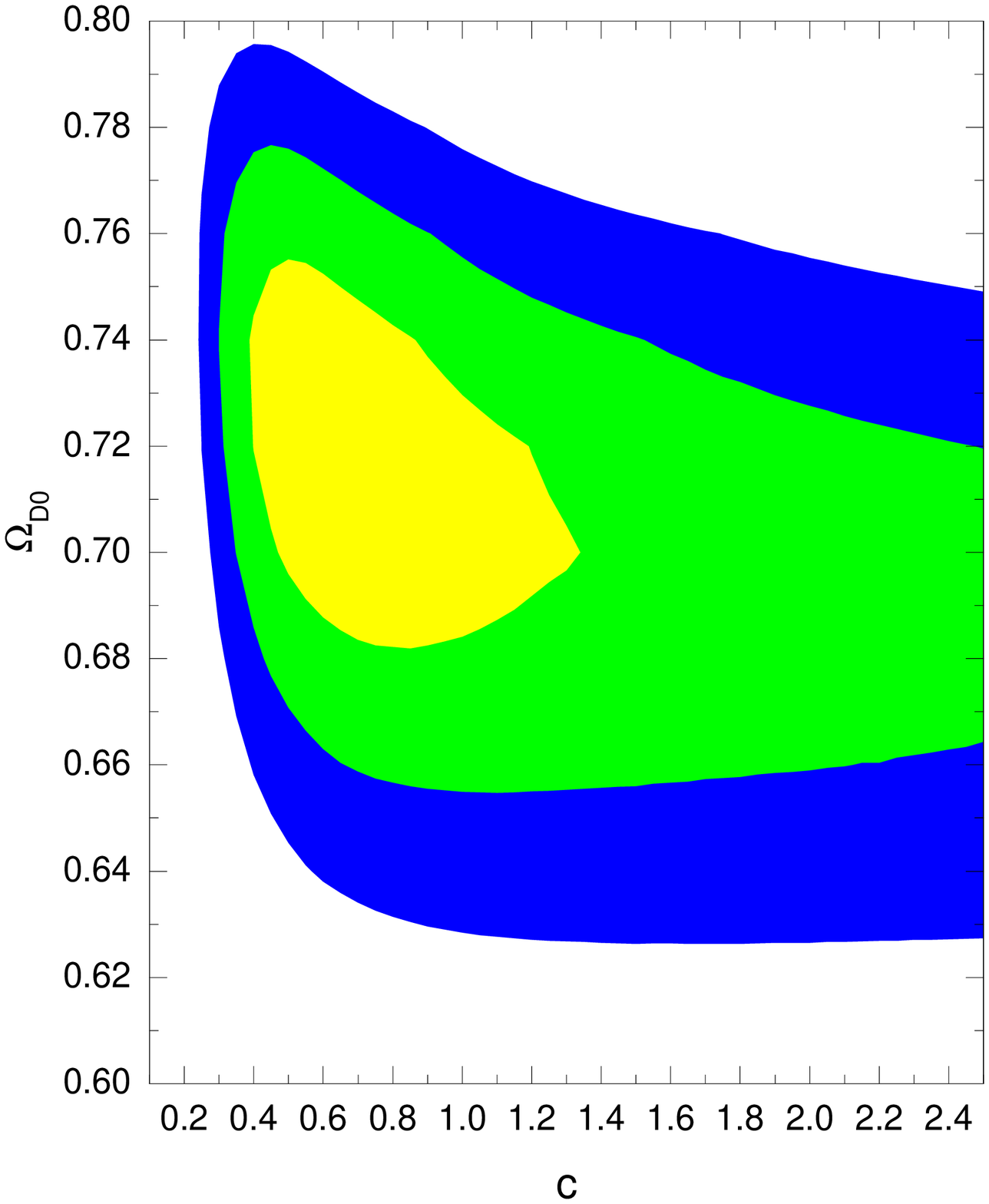}
    \end{minipage}}%
  \subfigure[]{
    \label{fig:mini:subfig:b} 
    \begin{minipage}[b]{0.5\textwidth}
      \centering
      \includegraphics[width=8cm,height=8cm]{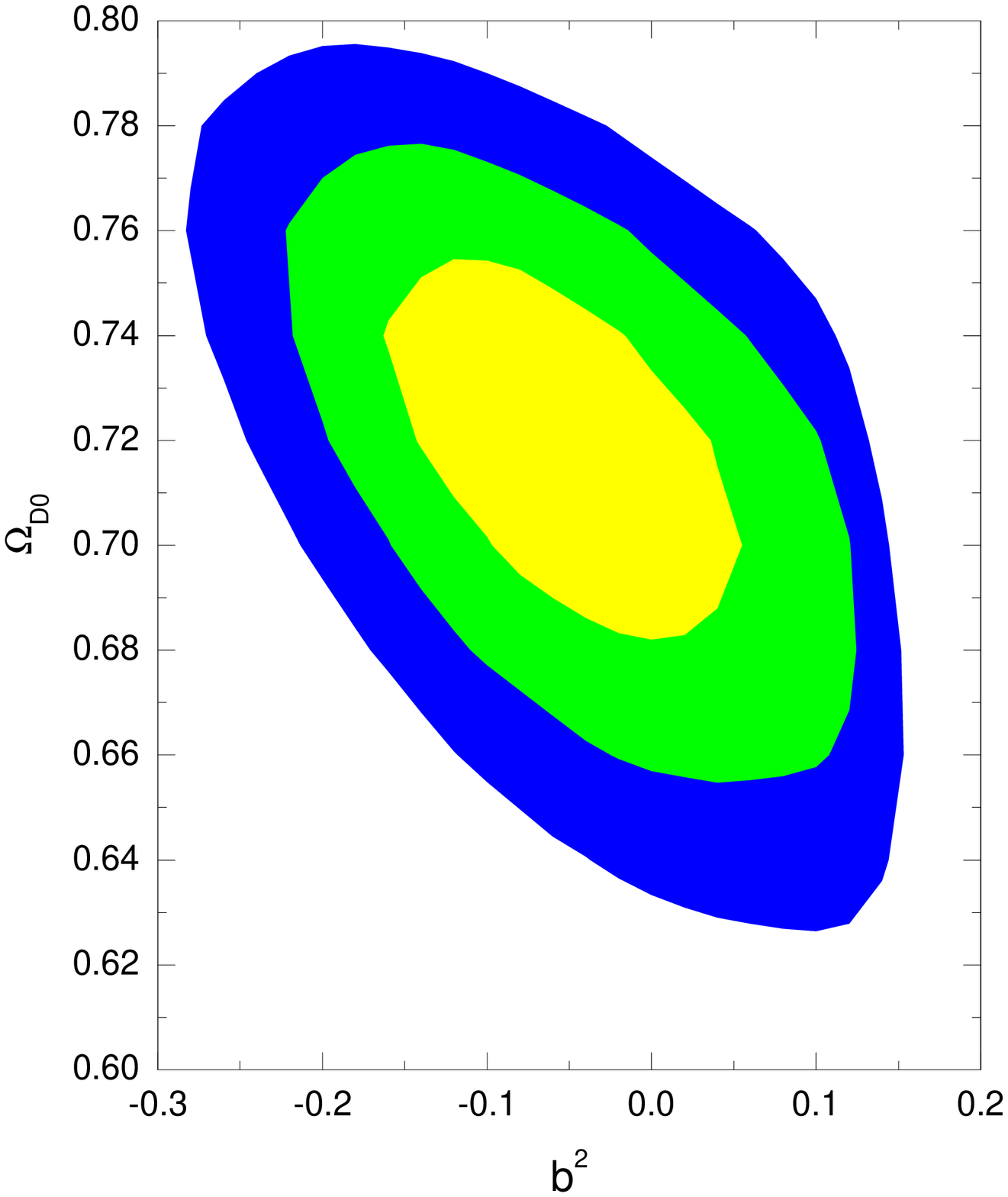}
    \end{minipage}}
  \caption{$(a)$The contours from the combination of
SN Ia, BAO, Lookback time in the interacting holographic dark energy
model for $c$ and $\Omega_{D0}$ at $1\sigma$, $2\sigma$, $3\sigma$
confidence level with $b^2=-0.059$. $(b)$The contours from the
combination of SN Ia, BAO, Lookback time for $b^2$ and $\Omega_{D0}$
at $1\sigma$, $2\sigma$, $3\sigma$ confidence level with $c=0.62$.}
  \label{fig:mini:subfig} 
\end{figure}

\begin{figure}
    \subfigure[]{
    \label{fig:mini:subfig:a} 
    \begin{minipage}[b]{0.5\textwidth}
      \centering
      \includegraphics[width=8cm,height=8cm]{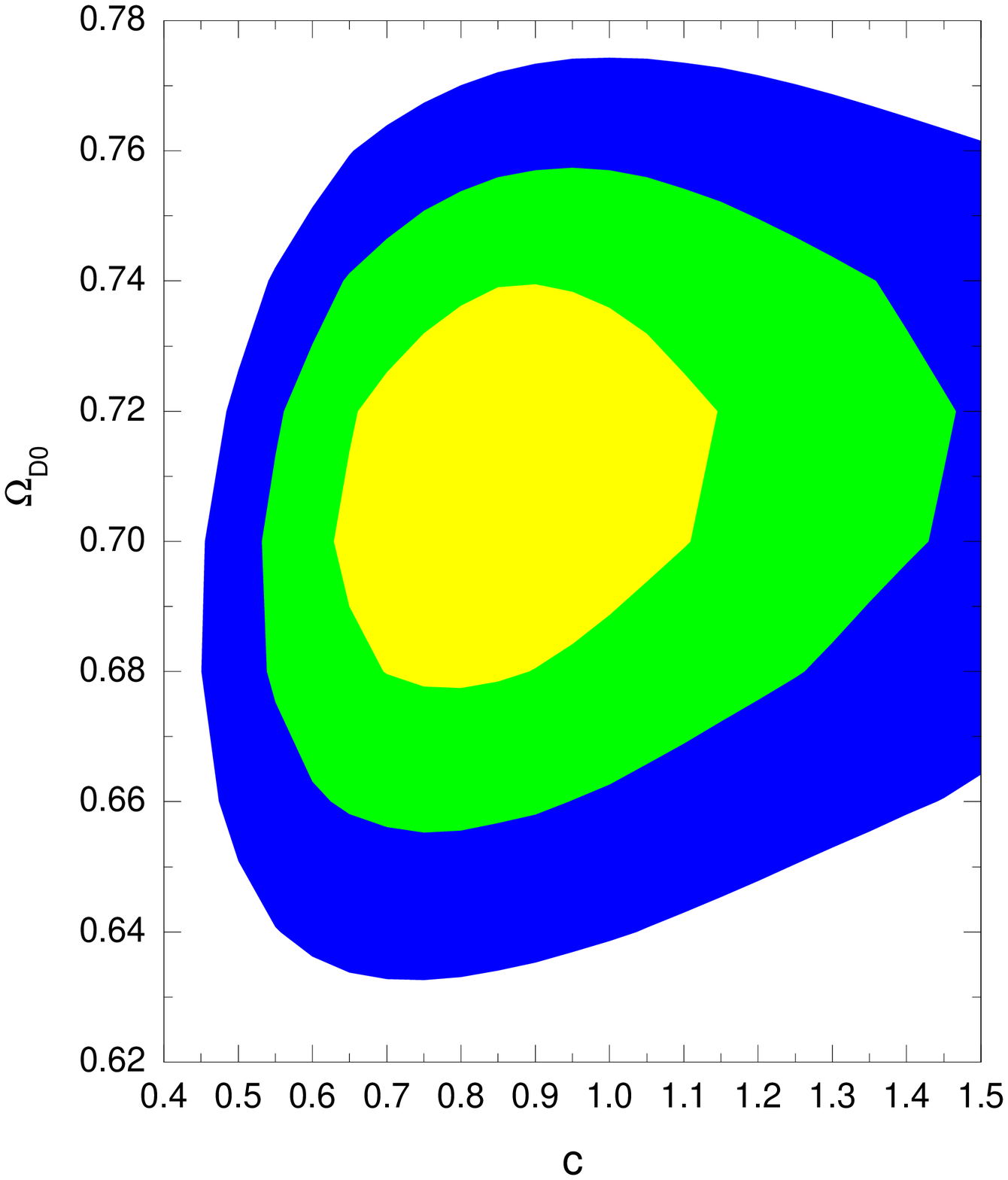}
    \end{minipage}}%
  \subfigure[]{
    \label{fig:mini:subfig:b} 
    \begin{minipage}[b]{0.5\textwidth}
      \centering
      \includegraphics[width=8cm,height=8cm]{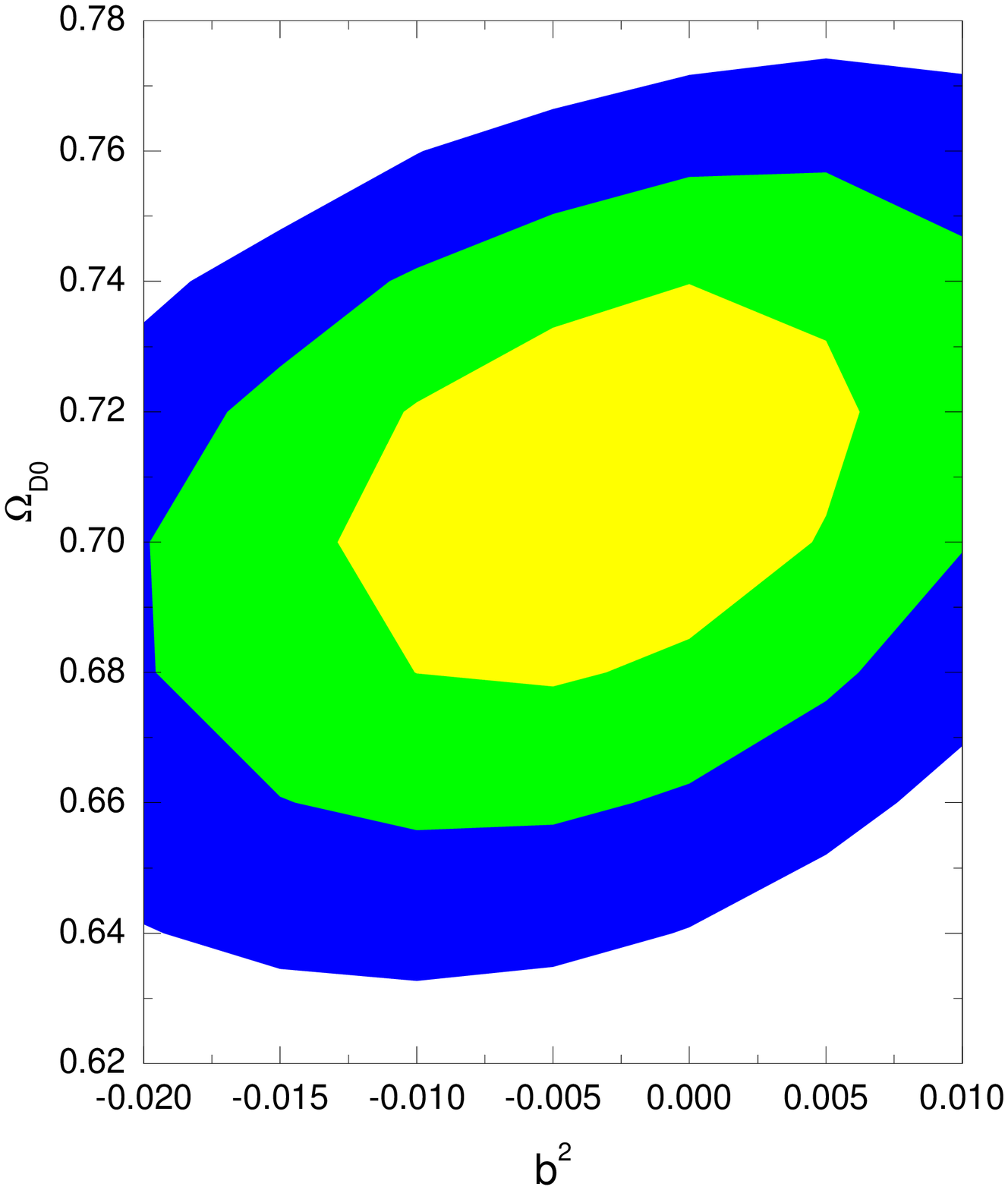}
    \end{minipage}}
  \caption{$(a)$The contours from the combination of
SN Ia, BAO, Lookback time, CMB in the interacting holographic dark
energy model for $c$ and $\Omega_{D0}$ at $1\sigma$, $2\sigma$,
$3\sigma$ confidence level with $b^2=-0.003$. $(b)$The contours from
the combination of SN Ia, BAO, Lookback time, CMB for $b^2$ and
$\Omega_{D0}$ at $1\sigma$, $2\sigma$, $3\sigma$ confidence level
with $c=0.83$.}
  \label{fig:mini:subfig} 
\end{figure}

The up-to-date gold SN Ia sample was compiled by Riess et al
\cite{10}. This sample consists of 182 data, in which 16 points with
$0.46<z<1.39$ were obtained recently by the Hubble Space Telescope
(HST), 47 points with $0.25<z<0.96$ by the first year Supernova
Legacy Survey (SNLS) and the remaining 119 points are old data. The
SN Ia observation gives the distance modulus of a SN at the redshift
$z$. The distance modulus is defined as
\begin{equation}
\mu_{th}(z;\textbf{P},\tilde{M})=5\log_{10}(d_L(z)/{\rm Mpc})+25=
5\log_{10}[(1+z)\int_0^z\frac{dz^{\prime}}{E(z^{\prime})}]+25-5\log_{10}H_0,
\end{equation}
where the luminosity distance
$d_L(z)=\frac{c(1+z)}{H_0}\int_0^z\frac{dz^{\prime}}{E(z^{\prime})}$,
the nuisance parameter $\tilde{M}=5\log_{10}H_0$ is marginalized
over by assuming a flat prior $P(H_0)=1$ on $H_0$,
$\textbf{P}\equiv{\{c,\Omega_D,b^2\}}$ describes a set of parameters
characterizing the given model. In order to place constraints on the
interacting holographic dark energy model, we perform $\chi^2$
statistics for the model parameter $\textbf{P}$
\begin{equation}
\chi_{SN}^2(\textbf{P},\tilde{M})=\sum_i{\frac{[\mu_{obs}(z_i)-\mu_{th}(z_i;\textbf{P},\tilde{M})]^2}{\sigma_i^2}}.
\end{equation}

Our analysis shows that if we use the SN Ia data, the constraint is
not good,and the $1\sigma$ range is rather large.

\begin{figure}
  \subfigure[1 $\sigma$ range of $q(z)$]{
    \label{fig:mini:subfig:a} 
    \begin{minipage}[b]{0.5\textwidth}
      \centering
      \includegraphics[width=8cm,height=8cm]{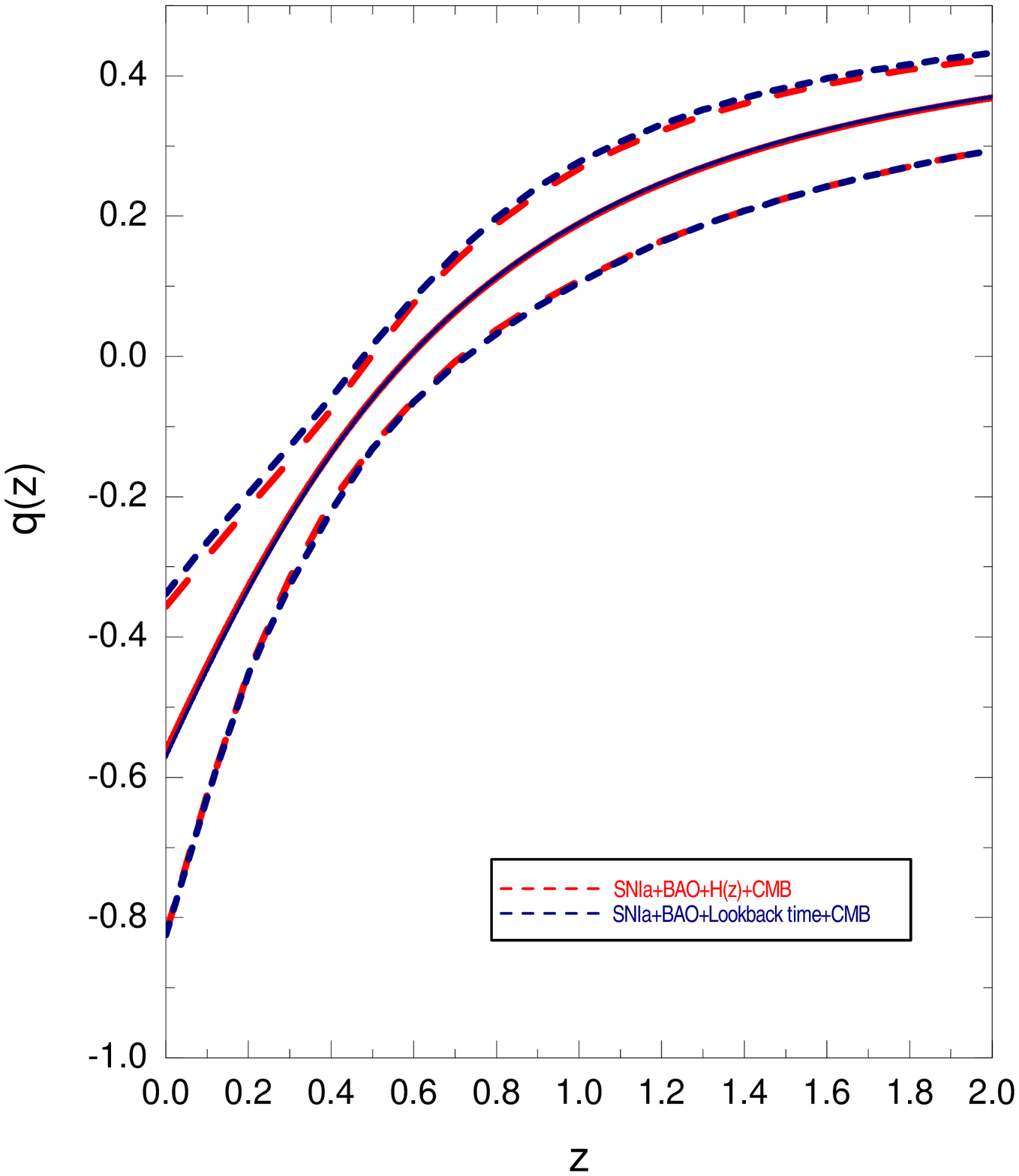}
    \end{minipage}}%
  \subfigure[1 $\sigma$ range of $w(z)$]{
    \label{fig:mini:subfig:b} 
    \begin{minipage}[b]{0.5\textwidth}
      \centering
      \includegraphics[width=8cm,height=8cm]{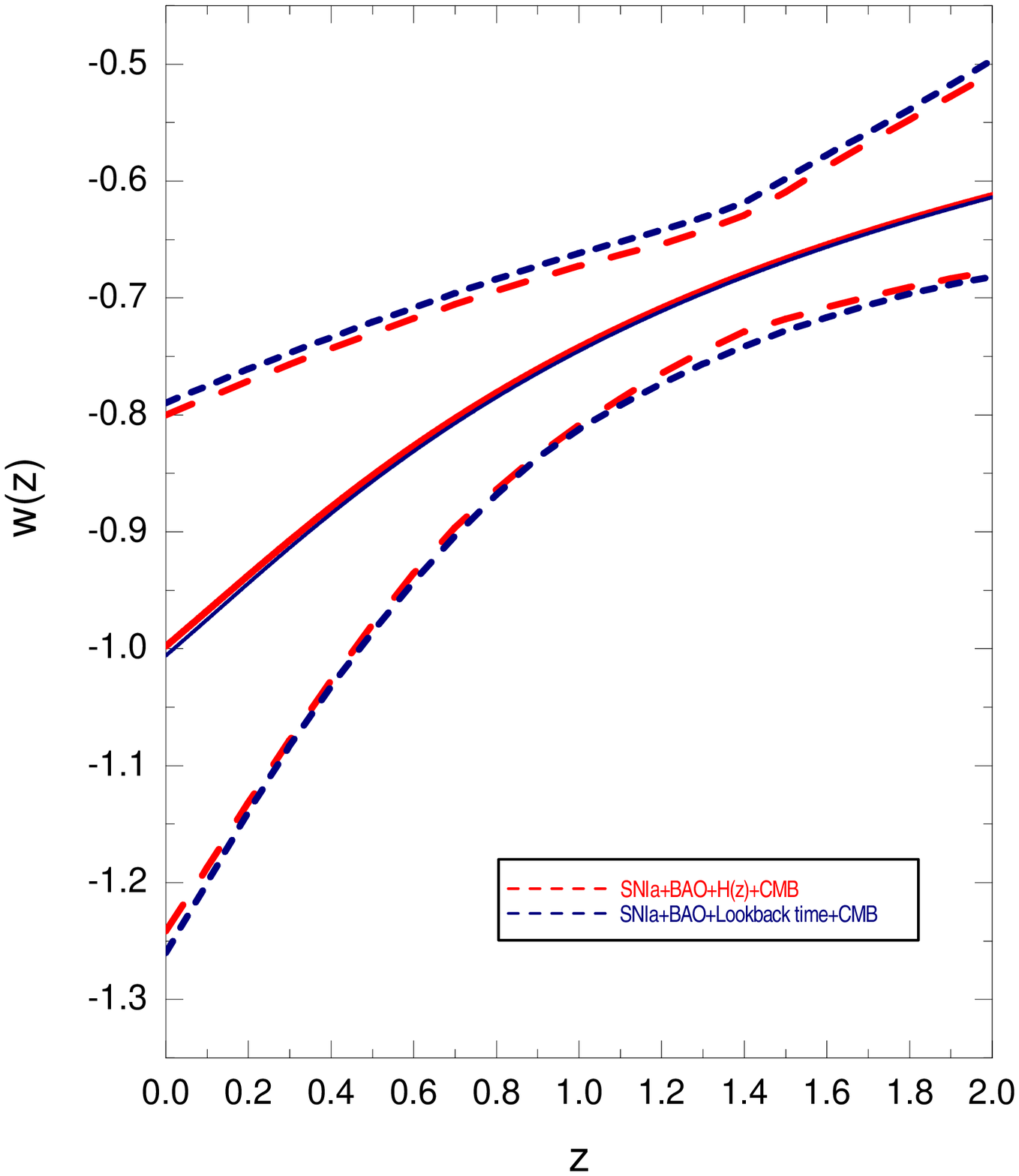}
    \end{minipage}}
  \caption{$(a)$The evolution of $q(z)$ within the 1 $\sigma$ range. $(b)$The evolution of $w(z)$ within the 1 $\sigma$ range. The 1 $\sigma$ range of each combination is between the same colored dash lines. And the solid lines are the best-fit curves of each combinations.}
  \label{fig:mini:subfig} 
\end{figure}

An efficient way to reduce the degeneracies of the cosmological
parameters is to use the SN Ia data in combination with the BAO
measurement from SDSS \cite{12} and the CMB shift parameter
\cite{11}. The acoustic signatures in the large scale clustering of
galaxies yield additional test for cosmology. Using a large sample
of 46748 luminous, red galaxies covering 3816 square degrees out to
a redshift of $z=0.47$ from the SDSS, Einstein et al \cite{12} have
found the model independent BAO measurement which is described by
the $A$ parameter
\begin{equation}
A=\sqrt{\Omega_m}E(z_{BAO})^{-1/3}[\frac{1}{z_{BAO}}\int_0^{z_{BAO}}\frac{dz^{\prime}}{E(z^{\prime})}]^{2/3}\\
=0.469(\frac{n_s}{0.98})^{-0.35}\pm0.017,
\end{equation}
where $n_s$ can be taken as $0.95$ \cite{WMAP3y} and $z_{BAO}=0.35$.
In our analysis we first investigated the joint statistics with the
SN Ia data and the BAO measurement. The result is shown in Figure 1,
where we show the contours of $68.3\%$ , $95.4\%$ and $99.7\%$
confidence levels. The fitted parameters with the $1\sigma$
errors are shown in Table 1.\\

\begin{figure}
  \subfigure[]{
    \label{fig:mini:subfig:a} 
    \begin{minipage}[b]{0.5\textwidth}
      \centering
      \includegraphics[width=8cm,height=8cm]{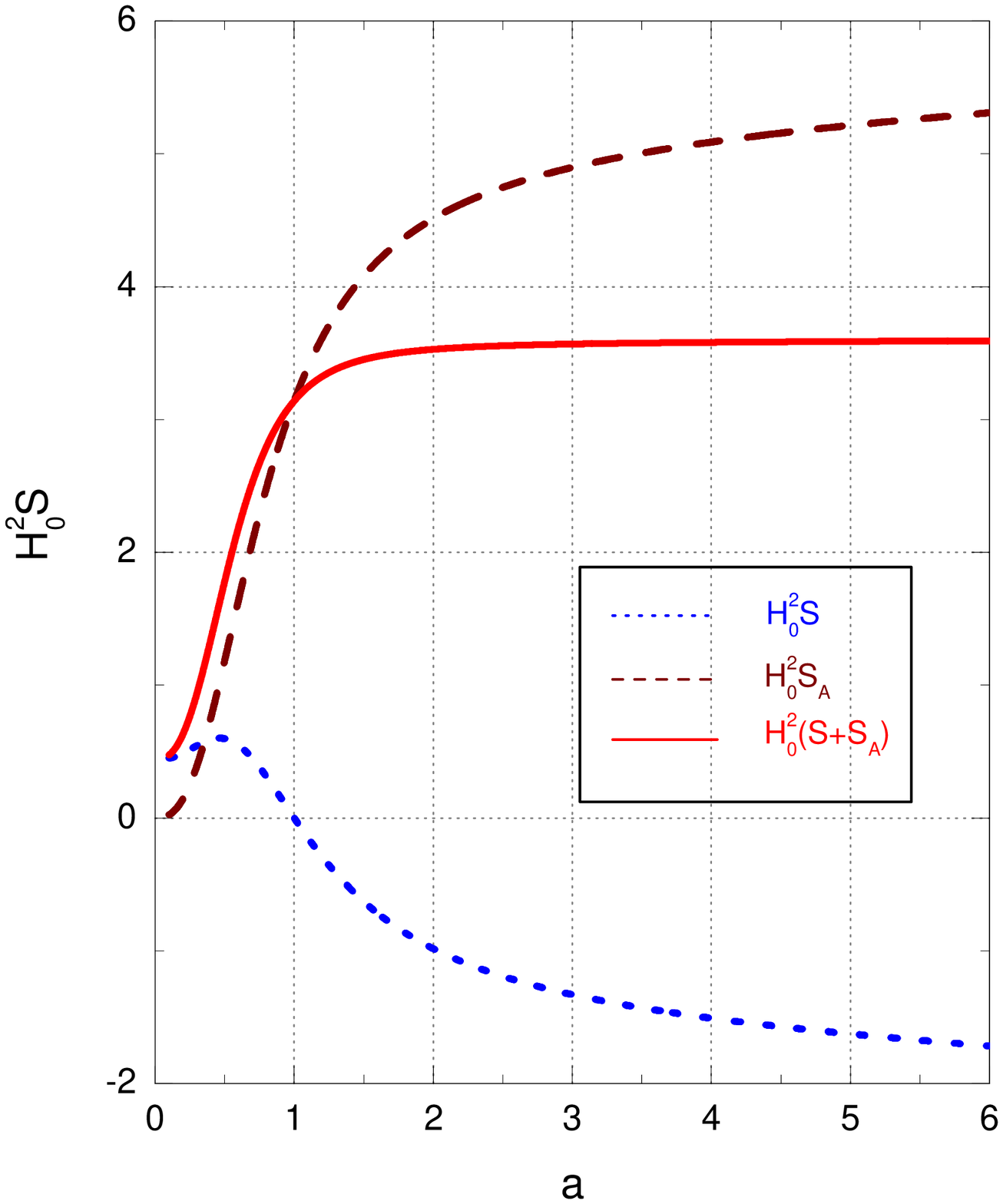}
    \end{minipage}}%
  \subfigure[]{
    \label{fig:mini:subfig:b} 
    \begin{minipage}[b]{0.5\textwidth}
      \centering
      \includegraphics[width=8cm,height=8cm]{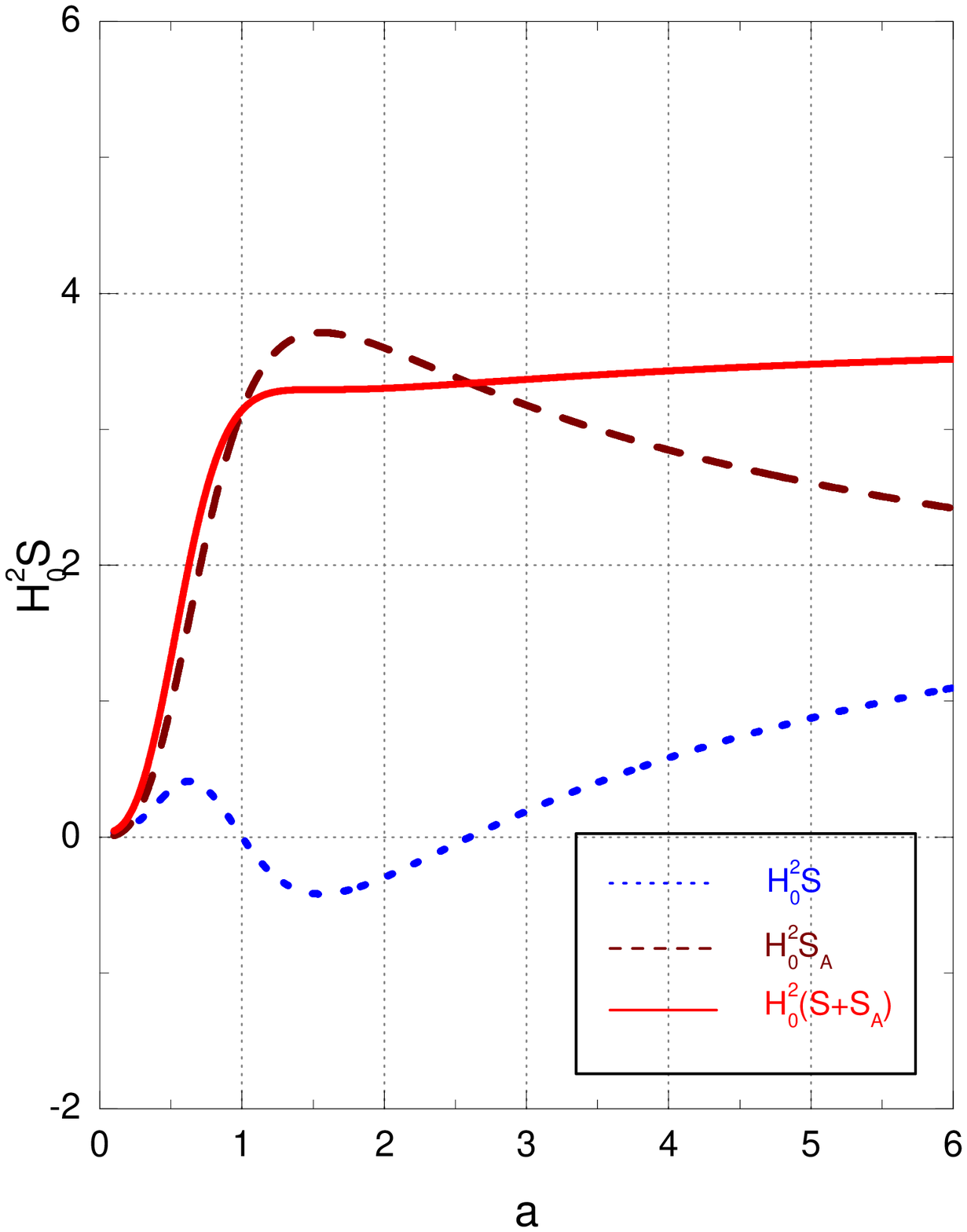}
    \end{minipage}}
  \caption{$(a)$The evolution of entropies with the $b^2=0.08$ and $c=1$ and the initial conditions $\Omega_{D0}=0.7$ and $H^2_0S_0=10^{-30}$.
  $(b)$The evolution of entropies with the best fit parameters of the combination Lookbacktime+SN Ia+BAO+CMB, $b^2=-0.003$ and $c=0.83$ and the initial conditions $\Omega_{D0}=0.71$ and $H^2_0S_0=10^{-30}$.}
  \label{fig:mini:subfig} 
\end{figure}

We also use the CMB shift parameter given by
\begin{equation}
R=\sqrt{\Omega_m}\int_0^{z_{ls}}\frac{dz^{\prime}}{E(z^{\prime})},
\end{equation}
where $z_{ls}=1089$. This CMB shift parameter $R$ captures how the
$l$-space positions of the acoustic peaks in the angular power
spectrum shift. Its value is expected to be the least model
independent and can be extracted from the CMB data. The WMAP3 data
\cite{WMAP3y} gives $R=1.70\pm 0.03$ \cite{11}. Now we can combine
the SN Ia, WMAP3 and SDSS data to constrain the interacting
holographic model. Using the $\chi^2$ statistics, contours from the
joint constraints SN Ia+BAO+CMB are shown in Figure 2. Comparing
with Figure 1, we see that the errors have been reduced
significantly in the joint analysis. The $1\sigma$ range of the
model parameters are listed in Table 1 for comparison.

\begin{figure}
  \subfigure[ SN Ia + BAO ]{
    \label{fig:mini:subfig:a} 
    \begin{minipage}[b]{0.5\textwidth}
      \centering
      \includegraphics[width=6cm,height=6cm]{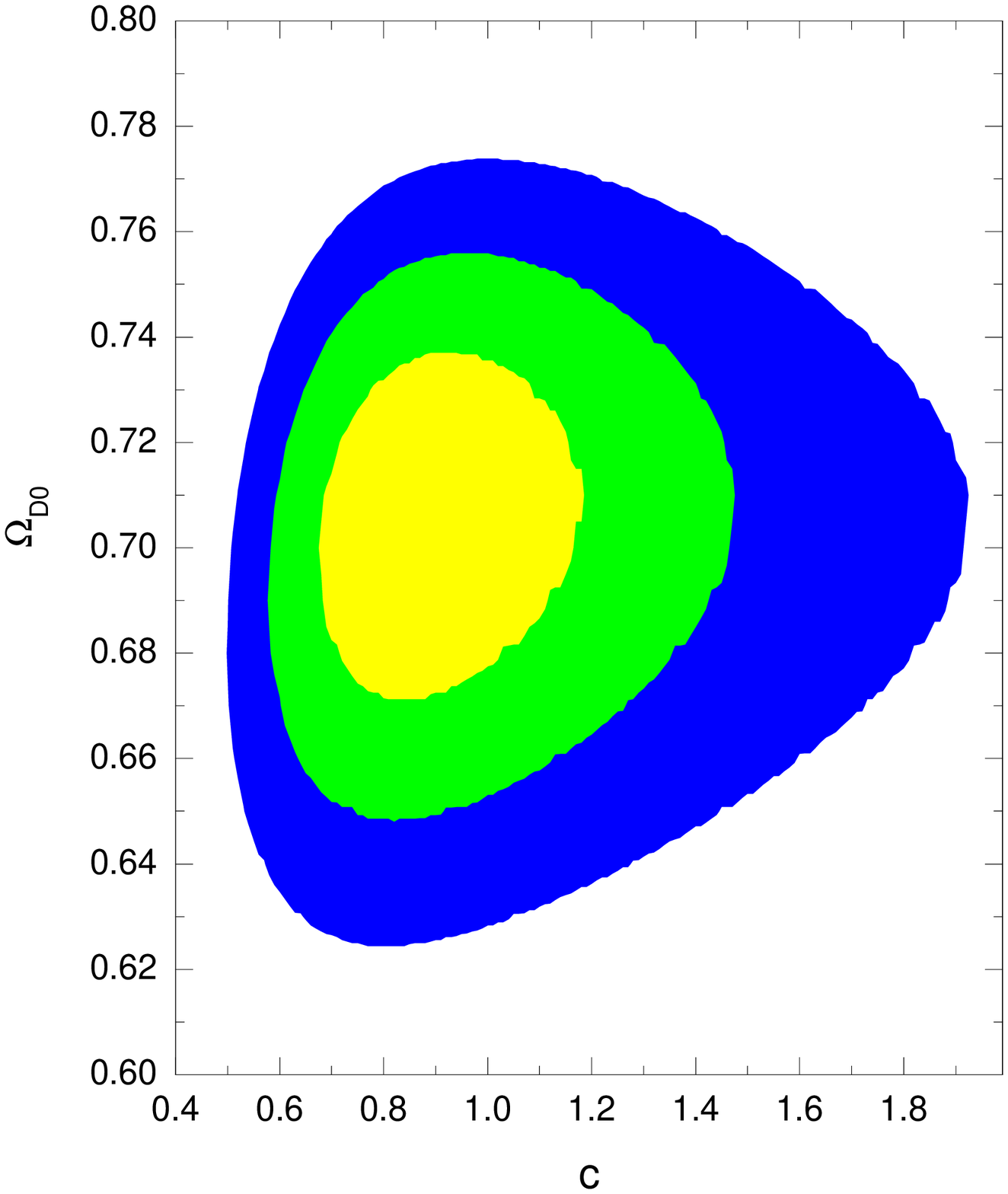}
    \end{minipage}}%
  \subfigure[SN Ia + BAO + CMB]{
    \label{fig:mini:subfig:b} 
    \begin{minipage}[b]{0.5\textwidth}
      \centering
      \includegraphics[width=6cm,height=6cm]{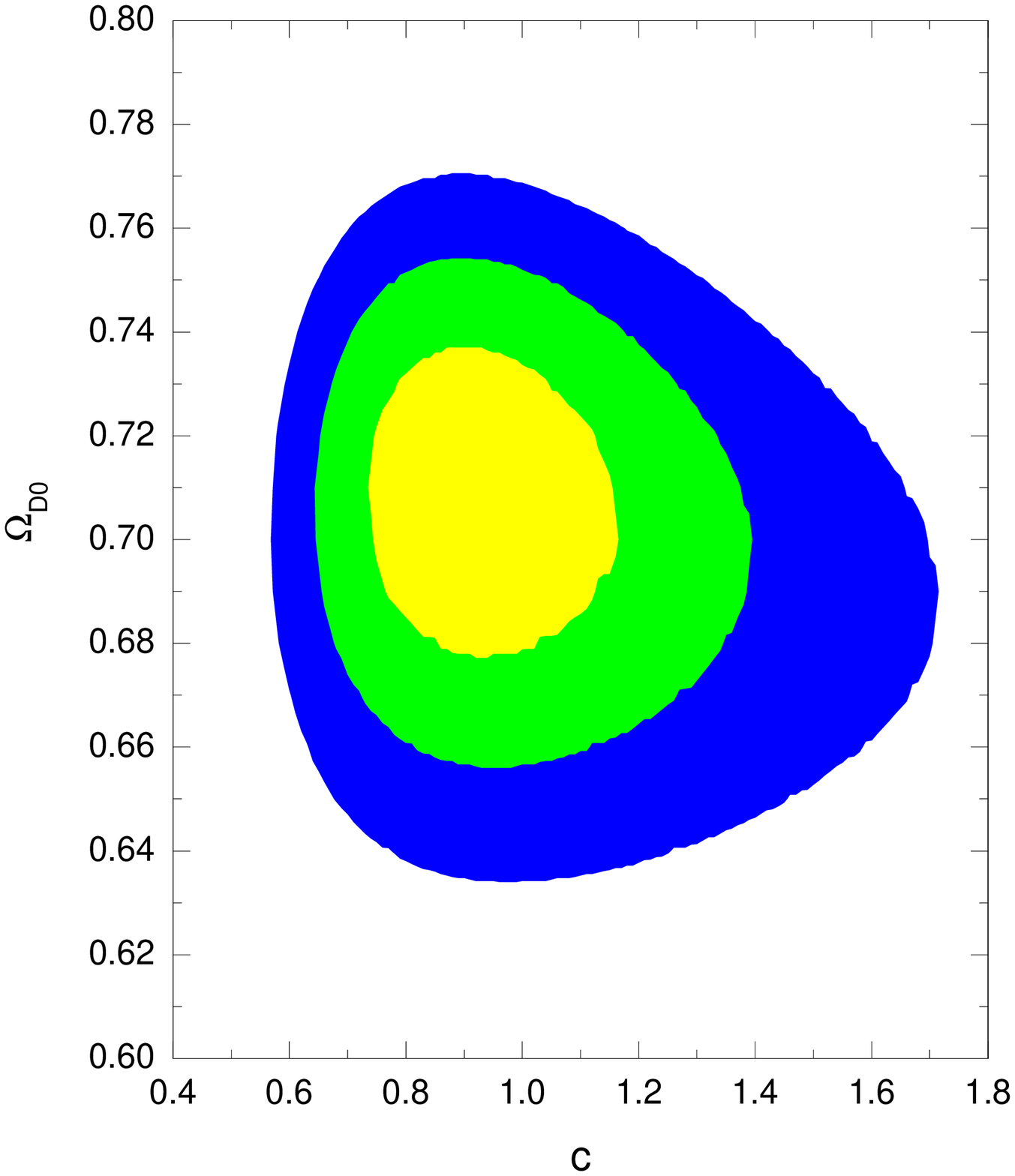}
    \end{minipage}}\\
    \subfigure[$H(z)$ + SN Ia + BAO ]{
    \label{fig:mini:subfig:a} 
    \begin{minipage}[b]{0.5\textwidth}
      \centering
      \includegraphics[width=6cm,height=6cm]{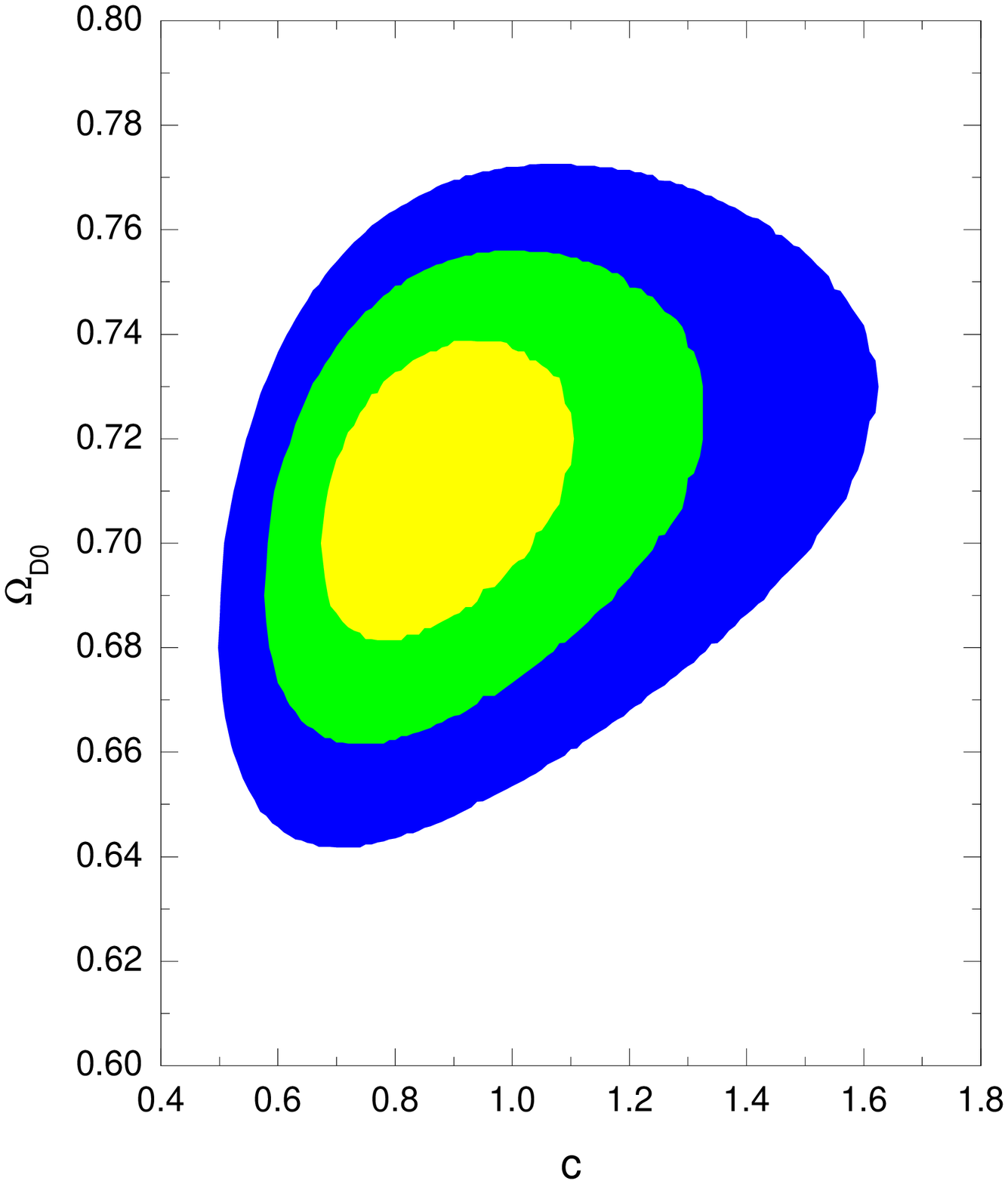}
    \end{minipage}}%
  \subfigure[Lookback time + SN Ia + BAO ]{
    \label{fig:mini:subfig:b} 
    \begin{minipage}[b]{0.5\textwidth}
      \centering
      \includegraphics[width=6cm,height=6cm]{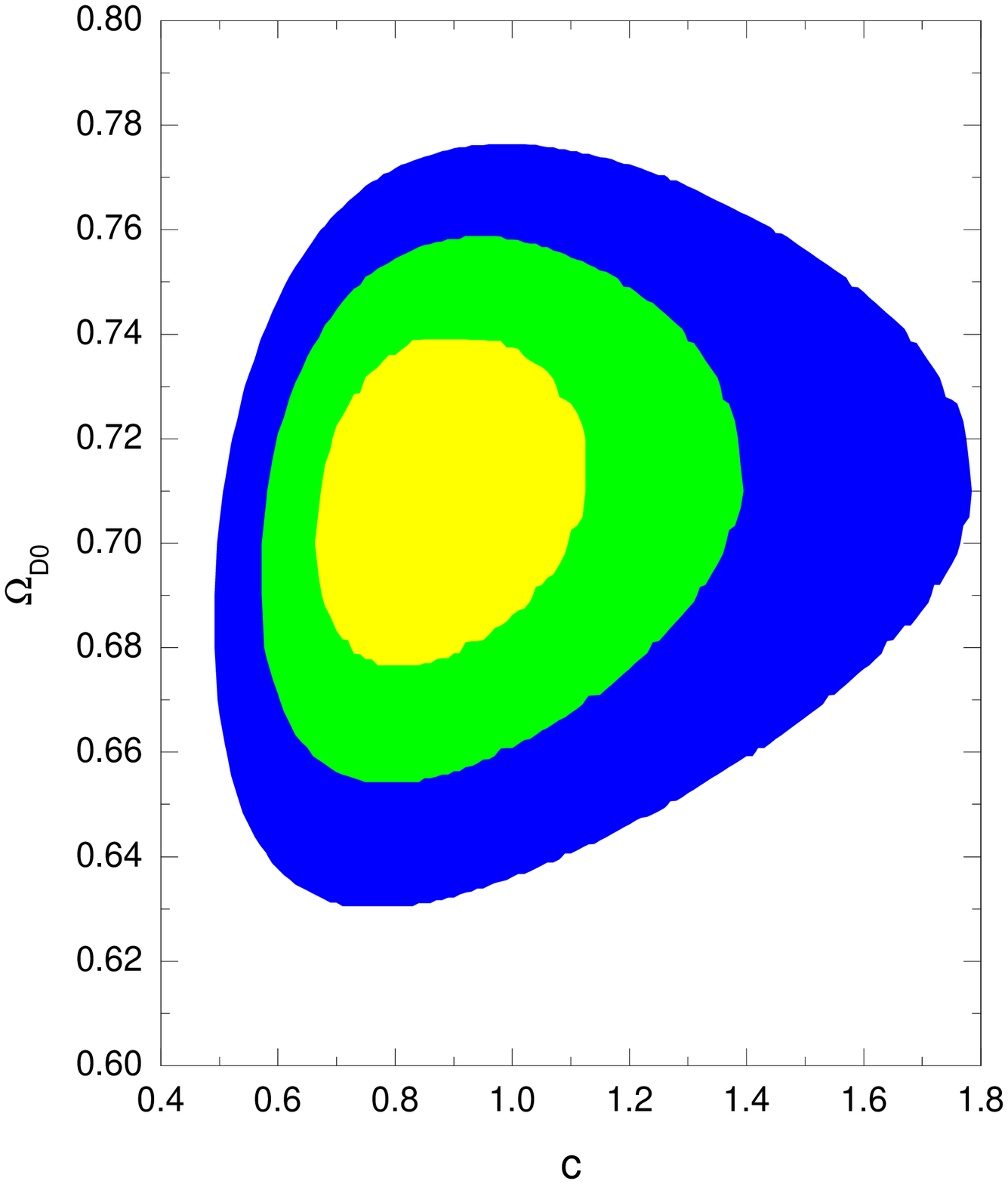}
    \end{minipage}}\\
    \subfigure[$H(z)$ + SN Ia + BAO +CMB]{
    \label{fig:mini:subfig:a} 
    \begin{minipage}[b]{0.5\textwidth}
      \centering
      \includegraphics[width=6cm,height=6cm]{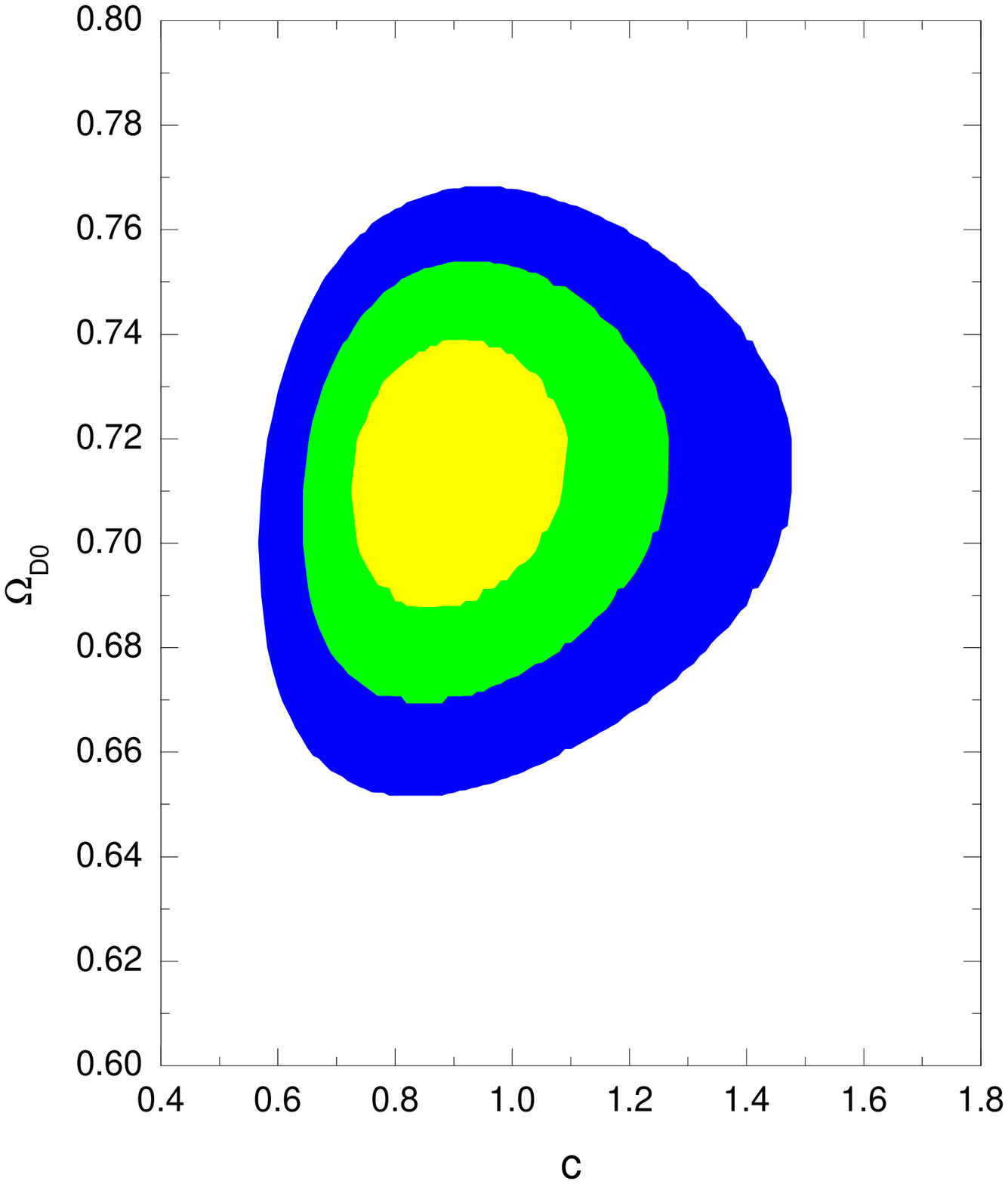}
    \end{minipage}}%
  \subfigure[Lookback time + SN Ia + BAO +CMB]{
    \label{fig:mini:subfig:b} 
    \begin{minipage}[b]{0.5\textwidth}
      \centering
      \includegraphics[width=6cm,height=6cm]{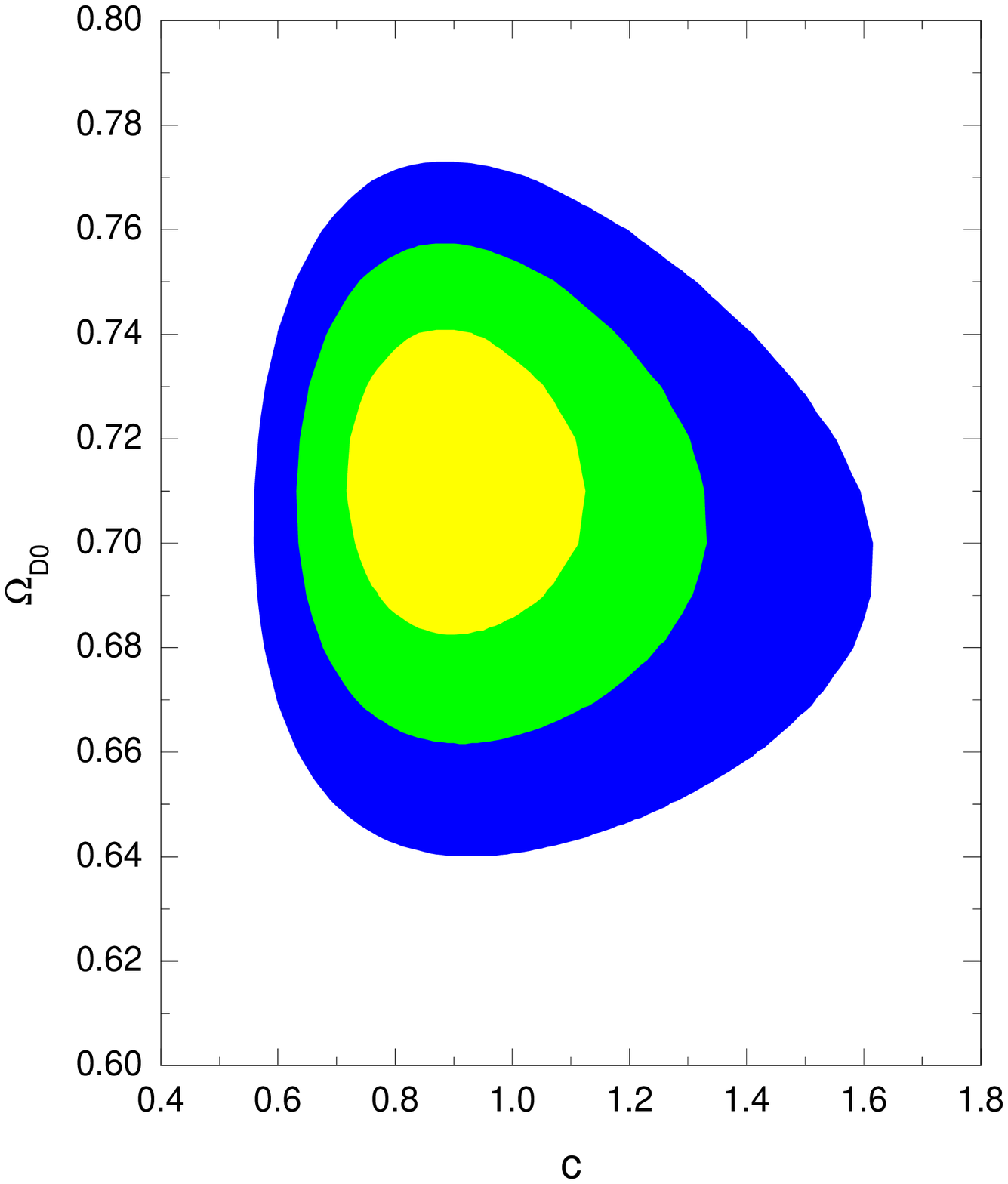}
    \end{minipage}}
  \caption{The contours in the holographic dark
energy model without interaction. This graph shows observational
contours in the $(c-\Omega_D)$ plane.}
  \label{fig:mini:subfig} 
\end{figure}

It is of interest to include the Hubble parameter data to constrain
our model. The Hubble parameter depends on the differential age of
the universe in terms of the redshift. In contrast to standard
candle luminosity distances, the Hubble parameter is not integrated
over. It persists fine structure which is highly degenerated in the
luminosity distance \cite{4}. Observed values of $H(z)$ can be used
to place constraints on the models of the expansion history of the
universe by minimizing the quantity
\begin{equation}
\chi_{h}^2(\textbf{P})=\sum_i{\frac{[H_{obs}(z_i)-H_{th}(z_i;\textbf{P})]^2}{\sigma_i^2}}.
\end{equation}

This test has been used to constrain several cosmological models
\cite{3,4}. However this test on its own cannot provide tight
constraint on the model. It is interesting to combine the $H(z)$
data with the data above to obtain tighter constraints on the
interacting holographic dark energy model. The result on the joint
analysis $H(z)$+SN Ia+BAO is shown in Figure 3 and the $1\sigma$
ranges of different parameters are listed in Table I. Because the
sample of $H(z)$ data is too small at this moment, the constraint on
the model by including $H(z)$ data is not very tight. We hope that
the future observations can offer more data of $H(z)$ so that
$\chi^2$ can be reduced. Adding the CMB shift parameter data, we
have shown the combined analysis $H(z)$+SN Ia+BAO+CMB shift in
Figure 4. Comparing with Figure 3, it is interesting to notice that
errors of model parameters have been significantly reduced.

\begin{table}
\scriptsize \caption{The best-fit data of the interacting
holographic dark energy model.}
\begin{center}
{
\begin{tabular}{|c|ccc|c|}
\hline
&c&$\Omega_{D0}$&$b^2$&$\chi_{min}^2$\\
\hline
$\mathrm{SN Ia+BAO}$&$0.53_{-0.22}^{+0.61}$&$0.72_{-0.04}^{+0.05}$&$-0.10_{-0.125}^{+0.131}$&156.24\\
\hline
$\mathrm{SN Ia+BAO+CMB}$&$0.84_{-0.25}^{+0.46}$&$0.70_{-0.04}^{+0.04}$&$-0.004_{-0.012}^{+0.012}$&158.45\\
\hline
$\mathrm{H(z)+SN Ia+BAO}$&$0.82_{-0.31}^{+0.89}$&$0.71_{-0.04}^{+0.05}$&$-0.005_{-0.075}^{+0.075}$&167.74\\
\hline
$\mathrm{Lookback time+SN Ia+BAO}$&$0.62_{-0.28}^{+1.22}$&$0.72_{-0.05}^{+0.05}$&$-0.059_{-0.126}^{+0.148}$&159.48\\
\hline
$\mathrm{H(z)+SN Ia+BAO+CMB}$&$0.84_{-0.25}^{+0.40}$&$0.71_{-0.04}^{+0.04}$&$-0.003_{-0.012}^{+0.010}$&167.75\\
\hline
$\mathrm{Lookback time+SN Ia+BAO+CMB}$&$0.83_{-0.25}^{+0.43}$&$0.71_{-0.04}^{+0.04}$&$-0.003_{-0.013}^{+0.012}$&160.08\\
\hline
\end{tabular}}
\end{center}
\end{table}

\begin{table}
\scriptsize \caption{The best-fit data of the noninteracting
holographic dark energy model.}
\begin{center}
{
\begin{tabular}{|c|cc|c|}
\hline
&c&$\Omega_{D0}$&$\chi_{min}^2$\\
\hline
$\mathrm{SN Ia+BAO}$&$0.88_{-0.20}^{+0.30}$&$0.71_{-0.03}^{+0.02}$&158.54\\
\hline
$\mathrm{SN Ia+BAO+CMB}$&$0.91_{-0.17}^{+0.25}$&$0.71_{-0.03}^{+0.02}$&158.64\\
\hline
$\mathrm{H(z)+SN Ia+BAO}$&$0.85_{-0.18}^{+0.26}$&$0.71_{-0.02}^{+0.02}$&167.77\\
\hline
$\mathrm{Lookback time+SN Ia+BAO}$&$0.85_{-0.18}^{+0.28}$&$0.71_{-0.03}^{+0.03}$&160.14\\
\hline
$\mathrm{H(z)+SN Ia+BAO+CMB}$&$0.88_{-0.15}^{+0.21}$&$0.71_{-0.02}^{+0.02}$&167.96\\
\hline
$\mathrm{Lookback time+SN Ia+BAO+CMB}$&$0.89_{-0.17}^{+0.23}$&$0.71_{-0.02}^{+0.03}$&160.32\\
\hline
\end{tabular}}
\end{center}
\end{table}

\begin{table}
\scriptsize \caption{The best-fit data of $\mathrm{\Lambda CDM}$ .}
\begin{center}
{
\begin{tabular}{|c|c|c|}
\hline
&$\Omega_{m0}$&$\chi_{min}^2$\\
\hline
$\mathrm{SN Ia+BAO}$&$0.30_{-0.02}^{+0.02}$&160.18\\
\hline
$\mathrm{SN Ia+BAO+CMB}$&$0.29_{-0.02}^{+0.02}$&161.85\\
\hline
$\mathrm{H(z)+SN Ia+BAO}$&$0.30_{-0.02}^{+0.02}$&169.22\\
\hline
$\mathrm{Lookback time+SN Ia+BAO}$&$0.30_{-0.02}^{+0.02}$&161.54\\
\hline
$\mathrm{H(z)+SN Ia+BAO+CMB}$&$0.29_{-0.02}^{+0.02}$&170.99\\
\hline
$\mathrm{Lookback time+SN Ia+BAO+CMB}$&$0.29_{-0.02}^{+0.02}$&163.05\\
\hline
\end{tabular}
}
\end{center}
\end{table}

The constraint based on SN Ia data and the recently proposed
angular-redshift relation of compact radio sources are distance
based methods to probe cosmological models, now we are going to test
the model by using the time-dependent observable, the lookback time.
The new test is expected to provide a complementary test of the
model. This method has been employed in \cite{8,9,7,18}. The
lookback time -redshift relation is defined by
\begin{equation}
t_L(z;\textbf{P})=H_0^{-1}\int_0^z\frac{dz^{\prime}}{(1+z^{\prime})E(z^{\prime})},
\end{equation}
where $H_0^{-1}=9.78h^{-1}$ Gyr, and we use the present value of
$h=0.72$ given by the HST key project \cite{hst}, $\textbf{P}$
stands for the model parameters. To use the lookback time and the
age of the universe to test a given cosmological model, let's follow
\cite{8} to consider an object $i$ whose age $t_i(z)$ at redshift
$z$ is the difference between the age of the universe when it was
born at redshift $z_F$ and the universe age at $z$,
\begin{equation}
t_i(z)=H_0^{-1}[\int_{z_i}^{\infty}\frac{dz^{\prime}}{(1+z^{\prime})E(z^{\prime})}
-\int_{z_F}^{\infty}\frac{dz^{\prime}}{(1+z^{\prime})E(z^{\prime})}].
\end{equation}
Using the lookback time definition, we have
$t(z_i)=t_L(z_F)-t_L(z)$. Thus the lookback time to an object at
$z_i$ can be expressed as
\begin{equation}
t_L^{obs}(z_i)=t_L(z_F)-t(z_i)=[t_o^{obs}-t_i(z)]-[t_o^{obs}-t_L(z_F)]=t_o^{obs}-t_i(z)-df,
\end{equation}
where $df=t_o^{obs}-t_L(z_F)$ is the delay factor.

In order to estimate the parameters of our model, we minimize the
$\chi^2$ function
\begin{equation}
\chi_{age}^2(\textbf{P})=\sum_i{\frac{[t_L(z_i;\textbf{P})-t_L^{obs}(z_i)]^2}{\sigma_i^2+\sigma_{t_o^{obs}}^2}}+\frac{[t_o(\textbf{P})-t_o^{obs}]^2}{\sigma^2_{t_o^{obs}}},
\end{equation}
where $\sigma_i=1$ Gyr is the uncertainty in the individual lookback
time to the $i$th galaxy cluster of our sample and
$\sigma_{t_o^{obs}}=1.4$ Gyr stands for the uncertainty on the total
age of the universe until now. The current age of the universe in
our analysis is taken as 14.4 Gyr. The second term in the $\chi^2$
expression was introduced to make sure that the cosmological model
can estimate the age of the universe at present in addition to
describing the age of the universe at high redshift. Since the delay
factor $df$ does not appear explicitly in the theoretical value of
$t_L(z_i)$, we will treat it as a nuisance parameter and marginalize
it in our calculation. The joint statistical analysis of the
combined observations including lookback time+SN Ia+BAO has been
done and the result is shown in Figure 5. Comparing to the analysis
of SN Ia+BAO shown in Figure 1, we noticed that the parameter space
now is enlarged. This fact is expected and understood in term of the
conservative uncertainty assumed $(\sigma_i=1Gyr)$ for the
individual lookback time. In Figure 6, we have shown the combined
analysis including lookback time+SN Ia+BAO+CMB shift. It is easy to
see that adding the CMB shift data, the model parameters have been
constrained much tighter.

To illustrate the cosmological consequences led by the observational
constraints, we show the evolution cases of the equation of state
parameter $w(z)$ and the deceleration parameter $q(z)$ according to
the  best-fit values of our model parameters in Figure 7. It is easy
to see that our model can have the feature of $w$ crossing $-1$. Our
present equation of state and the deceleration parameter are
consistent with CMB data \cite{WMAP3y,7}.

It is interesting to notice that our best fit value of $b^2$, the
coupling between dark energy and dark matter, is negative. In the
holographic interacting dark energy model by employing the apparent
horizon as the IR cutoff \cite{19}, it was argued that an equation
of state of dark energy $w<0$ is necessarily accompanied by the
decay of the dark energy component into pressureless matter
$(b^2>0)$. However, in our model, negative $b^2$ can accommodate
reasonable equation of state of dark energy which is clearly shown
in the Figure 7. Another worry of the negative $b^2$ which implies a
transfer of energy from the matter to the dark energy is that it
might violate the second law of thermodynamics \cite{p}. In order to
check the second law of thermodynamics, we can employ the formula in
\cite{17}. Using the apparent horizon as a thermal boundary and
evaluating the entropy inside the apparent horizon from the Gibbs
law, we have shown the evolution of entropies in Figure 8. It is
easy to see that for the best fit negative $b^2$, entropy of matter
and fluids inside the apparent horizon plus the entropy of the
apparent horizon do not decrease with time. The generalized second
law of thermodynamics is still respected.

For the sake of comparison, we have shown the same contours of the
holographic dark energy model without interaction in Figure 9. And
the best-fit results are shown in Table II. Combined SN Ia+CMB+BAO
constraints on the holographic dark energy model without interaction
have been studied in \cite{20}. Here we have added the Hubble
diagram data which is not an integrated over effect and the
time-dependent observable analyses. Comparing with Figure 2, we find
that $2\sigma, 3\sigma$ confidence ranges in the $c-\Omega_{D0}$
plane are much smaller for the holographic dark energy model without
interaction. In the analysis we find that the difference between the
model with and without interaction is bigger when the model
parameter $c$ is bigger. We have also listed the best-fit results
for a flat $\mathrm{\Lambda CDM}$ model in Table III. At the first
glance, from the $\chi^2_{min}$, it gives us a sense that the
interaction between dark energy and dark matter gives a better
description of the combined observations although this interaction
is extremely small. Considering the additional degree of freedom, it
is still early to say that our interacting holographic dark energy
model is more favored than the $\mathrm{\Lambda CDM}$ model. However
one advantage of the interacting holographic model is that, unlike
the $\mathrm{\Lambda CDM}$ model, it can alleviate the coincidence
problem\cite{16,7}.

In summary, in this work we have performed a parameter estimation of
the interacting holographic dark energy model which could explain
the observed acceleration of our universe. We have analyzed data
coming from the most recent SN Ia samples, CMB shift, LSS
observation, $H(z)$ and lookback time measurements. Comparing with
the single observational test, we learnt that the joint analysis of
different observations based on different observables is powerful to
overcome the statistical uncertainties. We have got useful
consistent check of the interacting holographic dark energy model
and tighter constraints on the model parameters. The joint analysis
indicates that this is a viable model. In the $1\sigma$ range, it
can explain the transition of the equation of state from  $w>-1$ to
$w<-1$. It is worth noting that although the current $H(z)$ and
lookback time data do not provide very restrictive constraints,
richer samples of $H(z)$ data and more precise age measurements of
high-z objects will provide a complementary check of the cosmic
acceleration model. The joint statistical analysis is necessary to
test the model.

We observed that the best fit coupling between dark energy and dark
matter is negative, which indicates that there is a possible energy
transfer from the matter to the dark energy. Although the
generalized second law of thermodynamics is shown not threatened by
the best fit negative $b^2$, the holographic principle might still
violate in the future if there is a continuing energy transfer from
the matter to the dark energy, since in the late stage it is
possible to see that $S<S_A$ in Figure 8. In view of the unknown
nature of dark energy and dark matter, we can't say for certain the
direction of the energy transfer between dark energy and dark
matter. However from the observation data and the holographic
principle requirement, the nature thing we can learn is that in the
past there is an energy flow from the matter to the dark energy from
the observational data, while in the future there requires an energy
transfer from the dark energy to the matter to satisfy the
holographic principle. A nature description of the coupling between
dark energy and dark matter is called for, which is important to
influence the structure formation and the description of the
universe evolution.

\acknowledgments This work was partially supported by NNSF of
China, Ministry of Education of China and Shanghai Educational
Commission. Y.G.G. was supported by Baylor University, NNSFC under
Grants No. 10447008 and No. 10605042, CMEC under Grant No.
KJ060502, and SRF for ROCS, State Education Ministry. B. W. would
like to acknowledge helpful discussions with D. Pavon.

\end{document}